\RequirePackage{lineno} 

\documentclass[twocolumn,showpacs,preprintnumbers,amsmath,amssymb,nofootinbib]{revtex4}

\usepackage{graphicx}
\usepackage{dcolumn}
\usepackage{bm}
 
\usepackage{epstopdf}
\def\bea{\begin{eqnarray}}
\def\eea{\end{eqnarray}}

\def\pp{\mbox{$p$-$p$}}

\def\pa{\mbox{$p$-A}}
\def\auau{\mbox{Au-Au}}

\def\pbpb{\mbox{Pb-Pb}}
\def\ppb{\mbox{$p$-Pb}}
\def\aa{\mbox{A-A}}
\def\nn{\mbox{$N$-$N$}}

\def\pt{$p_t$}

\def\mpt{$\langle p_t \rangle$}
\def\nch{$n_{ch}$}

\begin{document} 

\setlength{\pdfpagewidth}{8.5in}
\setlength{\pdfpageheight}{11in}

\setpagewiselinenumbers
\modulolinenumbers[5]

\preprint{version 1.3}

\title{
Event-wise mean-$\bf p_t$ fluctuations vs minimum-bias jets (minijets) at the LHC
}

\author{Thomas A.\ Trainor}
\affiliation{CENPA 354290, University of Washington, Seattle, WA 98195}


\date{\today}
;
\begin{abstract}
Fluctuation measurements of event-wise mean transverse momentum \mpt\ for \pp\ and \pbpb\ collisions at the large hadron collider (LHC) have been reported recently. In that study it was concluded that the  strength of ``nonstatistical'' \mpt\ fluctuations decreases with increasing particle multiplicity \nch\ (or \aa\ centrality) and is nearly independent of collision energy over a large interval. Among several potential mechanisms for those trends onset of thermalization and collectivity are mentioned. The LHC analysis employed one fluctuation measure selected from several possibilities. An alternative fluctuation measure reveals strong increase of \pt\ fluctuations with \nch\ (or \aa\ centrality) and collision energy, consistent with previous measurements at the relativistic heavy ion collider (RHIC). The \pt\ fluctuation data for LHC \pp\ collisions can be described accurately by a two-component (soft+hard) model (TCM) in which the hard component represents dijet production. The data for \pbpb\ collisions are described accurately by a TCM reference for more-peripheral collisions (suggesting transparent collisions), but the data deviate quantitatively from the reference for more-central collisions suggesting modification of jet formation. Overall fluctuation data trends suggest that minimum-bias jets (minijets) dominate \pt\ fluctuations at both the LHC and RHIC.
\end{abstract}


\maketitle

 \section{Introduction}

Fluctuation measurements of {\em event-wise} mean transverse momentum, denoted in this study by \mpt, vs charge multiplicity $n_{ch}$ in \pp\ and \pbpb\ collisions at the large hadron collider (LHC) have been reported recently~\cite{aliceptfluct}. A principal motivation for such measurements is the search for evidence of the phase transition between a quark-gluon plasma (QGP) and a hadronic medium (hadron gas) in the form of excess fluctuations of a thermodynamic quantity, specifically event-wise  $\langle p_t \rangle$ as a proxy for a local temperature~\cite{stod}. The $\langle p_t \rangle$ fluctuation measure chosen from among several candidates is apparently based on the assumed thermodynamic context. The reported systematic behavior includes negligible energy dependence over a large energy interval and general decrease of fluctuation ``strength'' with increasing event multiplicity or centrality. Those results conflict sharply with previous relativistic heavy ion collider (RHIC) results employing alternative fluctuation measures in which \pt\ fluctuation amplitudes {\em increase strongly} with \auau\ collision centrality and with collision energy~\cite{ptfluct,ptscale,ptedep}.


It is reasonable to assume direct connections between underlying collision mechanisms and final-state collision structure in the form of yields, spectra and correlations. But how the structure should be characterized statistically and how results should be interpreted in terms of physical mechanisms is in question. 
The overarching goal should be consistent and interpretable descriptions of particle densities on a multidimensional momentum space varying event-wise over collision events and extraction of all information residing in such data distributions.

The present study focuses on fluctuation measurement. Fluctuations and angular correlations are intimately related~\cite{hijscale,ptscale}. The scale (bin size) dependence of fluctuations corresponds to the space distribution of angular correlations~\cite{inverse}. The more structured a multiparticle angular distribution the larger the event-wise fluctuations. Fluctuation measures should then be compatible with correlation measures, providing useful design constraints. 
 
 
Just as for Ref.~\cite{aliceptfluct} some previous fluctuation measurements were motivated by a search for excess or critical fluctuations near a QCD phase boundary~\cite{na49ptfluct,phenixptfluct,jeffptfluct}. Conjectured possibilities included (a) QGP formed in some special events comprising a small fraction of an event ensemble or (b) QGP formed in most events of some event class resulting in a general fluctuation excess relative to some statistical reference.   While (a) was not observed and evidence for (b) was marginal at the SPS~\cite{na49ptfluct}, excess mean-\pt\  fluctuations in at least most events were clearly apparent in the first \auau\ data from the RHIC~\cite{jeffptfluct}.

Subsequent measurements of \pt\ fluctuations, particularly 2D bin-size dependence, led to a surprising result: \pt\ angular correlations inferred from inversion of \pt\ fluctuation bin-size dependence revealed that {\em dijet production} is the dominant source of \pt\ fluctuations at RHIC energies~\cite{ptscale}. The collision-energy dependence of \pt\ fluctuations and inferred angular correlations is also consistent with QCD dijet production via parton scattering~\cite{ptscale}. Those results seemed consistent with {\em number} (distinguished from  \pt) angular correlations that similarly indicated a dominant role for dijet production~\cite{axialci,anomalous}.

In contrast, fluctuation measurements employing alternative statistical measures (per-pair rather than per-particle, those terms defined below) applied to RHIC data suggested a very different scenario: \pt\ fluctuations/correlations, {\em in ratio with ensemble-mean \pt}, decrease strongly with increasing \aa\ centrality (suggesting increased thermalization) and are essentially independent of collision energy~\cite{starvol}. The recent analysis of LHC \pp\ and \pbpb\ data~\cite{aliceptfluct} seems to confirm those results.


The present study is the followup to a recent analysis of ensemble-mean \pt\ (denoted here by $\bar p_t$) systematics~\cite{tommpt}, derived from LHC \pp, \ppb\ and \pbpb\ data~\cite{alicempt}, wherein the $\bar p_t$ data for three collision systems and several energies are accurately described by a two-component (soft+hard) model (TCM) featuring dijet production as the hard component predicted quantitatively by perturbative QCD (pQCD). In contrast, theory Monte Carlos based on final-state hadron or parton rescattering and/or transverse flows~\cite{ampt} fail to describe the $\bar p_t(n_{ch})$ data.

Resolution of the \pt\ fluctuation dichotomy requires detailed analysis of methods and interpretations.
I first review the definitions and properties of several statistical measures applied to event-wise \pt\ fluctuations, with reference to measure design criteria. I present  results from the LHC fluctuation analysis reported in Ref.~\cite{aliceptfluct} in alternative plotting formats, discussing the apparent physical implications of the various data trends. Finally, I relate the LHC results to RHIC results derived from \pt\ fluctuation inversion to \pt\ angular correlations and to measured dijet systematics. The evidence suggests that an existing comprehensive TCM scenario describing dijet manifestations in yields, spectra and correlations also describes \pt\ fluctuation data accurately and supports the conclusion that dijet production plays a central role in all aspects of hadron production in high-energy nuclear collisions.
 
This article is arranged as follows:
Section~\ref{lhcfluct} summarizes a recent study of \mpt\ fluctuations at the LHC.
Section~\ref{meths} introduces various methods for \pt\ fluctuation measurement.
Section~\ref{collmod} defines several basic models for \pp\ and \aa\ collisions.
Section~\ref{mpttcm}  defines a \pp\ TCM for ensemble-mean $\bar p_t$.
Section~\ref{alicempt}  summarizes LHC \pp\ \pt\ fluctuation data.
Section~\ref{pbpbfluct}  presents LHC \pbpb\ \pt\ fluctuation data compared to RHIC \auau\ data.
Section~\ref{scale}  relates \pt\ fluctuation measurements to \pt\ angular correlation measurements including dijet-related structure.
Sections~\ref{disc} and~\ref{summ} present discussion and summary,
and App.~\ref{aatcm} presents a $\bar p_t$ TCM for \aa\ collisions.

\section{LHC $\bf \langle p_t \rangle$ fluctuation analysis} \label{lhcfluct}

Reference~\cite{aliceptfluct} (ALICE collaboration)  reports measurements of \mpt\ fluctuations vs $n_{ch}$ in \pp\ and \pbpb\ collisions at the LHC.
 Here I summarize conclusions of that study and in following sections consider details of \pt\ fluctuation  trends. In Ref.~\cite{aliceptfluct} event-wise mean-\pt\  is denoted by $M_\text{EbE}(p_\text{T})$ whereas in  this study the symbol is $\langle p_t \rangle$.

 In the context of a {\em temperature narrative} the primary motivation for \mpt\ fluctuation analysis is a search for critical fluctuations of (local) temperature associated with a QCD phase boundary. A QCD phase transition or critical point may ``go along with'' critical fluctuations in a thermodynamic quantity such as temperature represented (within an assumed theoretical context) by $\langle p_t \rangle$. The adopted dimensionless ratio $\sqrt{C_m} / \bar p_{t,m}$ is said to ``quantify the strength of the non-statistical fluctuations in units of the [ensemble] average transverse momentum $M(p_T)_m$ [$= \bar p_{t,m}$] in the multiplicity class $m$.''


The basic fluctuation measure is the ratio $C = \bar B / \,\overline{n_{ch}(n_{ch} - 1)}$, where $\bar B$ is a variance difference defined below. The \pt\ acceptance is $p_t \in [0.15,2]$ GeV/c, where the lower limit is determined by the detector and the imposed upper limit may represent an effort to exclude dijet contributions (e.g. \pt\ spectra below 2 GeV/c have been interpreted entirely in terms of a thermalized flowing bulk medium with no jet contribution~\cite{nojets}). The lower limit has important consequences for ensemble-mean $\bar p_t$ and \pt\ fluctuation analysis. 
No jet contribution to \pp\ or \pbpb\ fluctuation data is acknowledged.

\pp\ fluctuation data reported at several LHC energies and corresponding to non-single-diffractive (NSD, low-multiplicity) conditions show no significant collision-energy dependence over a large energy interval. The \pp\ data are said to exhibit a clear power-law dependence on charge multiplicity \nch\  but deviate from a {\em linear-superposition} reference $\propto 1/n_{ch}^{0.5}$. Several theory Monte Carlos (MCs) show ``qualitative agreement'' with the \pp\ fluctuation data, quite different from the qualitative disagreements for $\bar p_t(n_{ch})$ data reported in Ref.~\cite{alicempt}.

Conjectured \pp\ $\langle p_t \rangle$ fluctuation mechanisms are resonance decays, jets and quantum correlations. To account for those ``conventional mechanisms'' in \pbpb\ collisions \pp\ results are assumed as a reference. \pp\ data provide a ``model-independent baseline....'' Nontrivial results in \aa\ would then be signaled by ``modification of the fluctuation pattern with respect to the \pp\ reference.'' 

``Fluctuations [of $\langle p_t \rangle$ in \pbpb] were found to decrease with collision centrality, as generally expected in a dilution scenario caused by superposition of partially independent particle-emitting sources....
Deviations from a simple superposition scenario have been reported.'' 
The linear-superposition reference (for $\sqrt{C} / \bar p_t$) is again assumed to be $\propto 1/n_{ch}^{0.5}$. 
Relative to the \pp\ power-law trend \pbpb\ peripheral data follow a similar power law, but more-central \pbpb\ data first rise sharply relative to the \pp\ trend and then fall off more slowly. That is noted as a remarkable correspondence given the major disagreements for $\bar p_t$ data as in Refs.~\cite{alicempt,tommpt}.

The \pbpb\ results are said to be consistent with string percolation or the onset of thermalization and collectivity. No critical behavior is evident. There is possibly evidence for initial-state density fluctuations. Several Monte Carlos are in ``qualitative agreement'' with \pbpb\ data.
HIJING, a model of \aa\ collisions, is said to follow a linear-superposition power-law reference and is inconsistent with the \pbpb\ fluctuation data.
Qualitative agreement is reported between \pbpb\ data and \auau\ data and with Monte Carlo models that incorporate collective phenomena. It is concluded there is no significant energy dependence of \auau\ or \pbpb\ fluctuation data.

\section{Statistical Analysis methods} \label{meths}

Different statistical measures applied to the same particle data may lead to contradictory physical interpretations. Does that mean statistical analysis is arbitrary, that collision mechanisms cannot be inferred from particle data? Resolution of  ambiguity requires detailed comparison of measures and the requirement that a valid description must confront all analysis results consistently. 

 I compare several analysis methods in the context of conventional statistics, including the central limit theorem and Pearson's correlation coefficient. 
To simplify algebraic relations I eliminate cumbersome summation notation  where possible and introduce a compact and self-consistent symbol set based on common usage. 

The {\em central limit theorem} (CLT) provides a basic reference for fluctuation analysis. The CLT asserts that for certain conditions---uncorrelated samples from a fixed parent process---certain moments of the sample population are invariant under scale transformations~\cite{clt,jeffptfluct}.

\subsection{Basic statistical quantities}

A random variable (RV) represents a set of samples from a parent process (e.g.\ density distribution or sequence of physical events such as nuclear collisions). I assume some detector angular acceptance $(\Delta \eta,\Delta \phi)$ on pseudorapidity $\eta$ and azimuth $\phi$ that may be partitioned into some number of bins $M$. The basic event-wise RVs for the present study are $P_t$ and $n_{ch}$ representing sums over those charged particles falling within an angular-acceptance bin in a collision event (the entire detector acceptance or some fraction thereof). We take $n_{ch} \rightarrow n$ below to lighten the notation. 
The ensemble-mean bin pair number is then
\bea \label{nnmin1}
 \overline{n(n-1)} &=& \frac{1}{N_{evt}} \sum_{k = 1}^{N_{evt}} n_{k}(n_{k}-1)
\\ \nonumber
&=& \bar n^2 + \bar n \Delta \sigma^2_n,
\eea
which defines the normalized number {\em variance difference} $\Delta \sigma^2_n = (\sigma^2_n - \bar n) / \bar n $ as a {\em per-particle} measure of number {\em fluctuation excess} relative to a statistical reference. For number fluctuations $ \sigma^2_{n,ref} =\bar n$ is the Poisson reference variance. Generally $\Delta \sigma^2_x \equiv (\sigma^2_x -  \sigma^2_{x,ref}) / \bar n$ is a per-particle measure of variance excess for RV $x$ (consistent with Pearson's correlation coefficient defined below).
A  compound RV such as $\langle p_t \rangle =  (\bar P_t + \delta P_t) / (\bar n + \delta n)$ (where $\delta x$ is an event-wise deviation from the ensemble mean) is a complex statistic with fluctuations represented by a series of variances and covariances of elementary RVs.

In the data analysis and discussion that follows I assume that multiplicities are integrated within one unit of $\eta$, so for instance $dn_{ch}/d\eta \approx n_{ch} / \Delta \eta  \rightarrow n_{ch}$, but in the figures the density ratios are made explicit. Multiplicity variables are event-wise random variables $n_x$ with means $\bar n_x$. To simplify notation I omit the bars on multiplicity variables unless there is ambiguity. 

\subsection{Variance-based fluctuation measures}

I assume that fluctuating RV $x$ follows a peaked distribution with mean value $\bar x$ and characteristic r.m.s.\ width $\sigma_x$ (linear dispersion measure). Variance $\sigma^2_x$ is a conventional fluctuation measure.
The single-particle \pt\ variance ($x \rightarrow p_t$) for a specific acceptance bin is
\bea
 \sigma^2_{p_t} &=& \frac{1}{\bar n} \overline{\left \{ \sum_{i=1}^{n} (p_{t,i} - \bar p_t)^2\right\}} .
\eea
The variance of bin-sum $P_t$ given event-ensemble means $\bar n$ and $\bar p_t$ is $ \sigma^2_{P_t} = \overline{(P_t - \bar n \bar p_t)^2}$ . The {\em conditional} variance of $P_t$ {\em given event-wise bin multiplicity $n$} is
\bea
 \sigma^2_{P_t|n} &=&  \overline{(P_t -  n \bar p_t)^2}
 \\ \nonumber
 &=&\sigma^2_{P_t} - 2 \bar p_t \sigma^2_{P_t n} + \bar p_t^2 \sigma^2_n,
\eea
where $\bar p_t = \bar P_t / \bar n$, and $ \sigma^2_{P_t n}$ measures the $n$-$P_t$ covariance that prompted the introduction of measure $\Phi_{p_t}$ (defined below) to study equilibration in \aa\ collisions~\cite{marek}.

The  ``correlator'' $C$~\cite{starvol,alicempt} defined below is a ratio of means. Its numerator  $\bar B$ can be re-expressed more simply as a {\em variance difference}
\bea \label{bbar}
B &=&   \sum_{i\neq j = 1}^{n,n-1}(p_{t,i} - \bar p_t)(p_{t,j} - \bar p_t)
 \\ \nonumber
 &=& (P_t -  n \bar p_t)^2 - n\langle(p_t - \bar p_t)^2\rangle
\\ \nonumber
\bar B &=&   \sigma^2_{P_t|n} - \bar n \sigma^2_{p_t},
\eea
which is zero for CLT conditions (stationary parent process and no significant two-particle correlations)~\cite{clt,jeffptfluct}.

$B$ has been described in terms of covariances~\cite{starvol,aliceptfluct}, but the usage is misleading. Whereas the structure of B expressed in the first line may suggest a covariance representing correlations (between what two quantities?) the reality is a variance difference (as expressed in the third line) representing a conditional {\em fluctuation excess} and describing a {\em single} RV. The algebraic relation between fluctuations and correlations was established in Ref.~\cite{inverse}.
$\bar B$ is a specific example (over a limiting scale interval) of total variance difference $\Delta \Sigma^2_{P_t|n}$ defined in Refs.~\cite{jeffptfluct,inverse,ptscale}.

\subsection{Pearson's correlation coefficient}

A covariance describes the relation between two distinct RVs, for instance event-wise sums from different acceptance bins. In some cases a covariance may be related to (normalized by) marginal variances of the individual RVs. The prototype is Pearson's correlation coefficient.
Pearson's normalized covariance or {\em  product-moment correlation coefficient} for a joint event distribution on two random variables $x_a$ and $x_b$ in separate bins $(a,b)$ is~\cite{pearson}
\bea
r_{ab} &=& \frac{\sigma^2_{ab}}{\sqrt{\sigma^2_a\, \sigma^2_b}} \in [-1,1],
\eea
the covariance $\sigma^2_{ab}$ normalized by the geometric mean of marginal (single-bin) variances as a normalization factor. Marginal variances $\sigma^2_x$ may represent actual marginal data projections, a mixed-event data reference or an idealization such as a Poisson reference assuming factorization of  the joint distribution.
The geometric mean in  the denominator implies factorization assuming a reference with no $a$-$b$ correlations (CLT conditions).

In the present context a bin-sum $P_t$ {\em covariance} between bins $a$ and $b$ can be defined as
\bea \label{bab}
\bar B_{ab} &=& \overline{(P_{t} - n \bar p_t)_a(P_{t} - n \bar p_t)_b} ,
\eea
and the pair-number reference becomes
\bea
 \overline{n(n-1)}  &\rightarrow& \overline{n_a \, n_b} \rightarrow \bar n_a \bar n_b
\eea
consistent with CLT conditions and factorization.
Pearson's  correlation coefficient for fluctuations of sum \pt\ in each of bins $a$ and $b$ given the bin multiplicities is then
\bea \label{rab}
r_{ab} &=& \frac{ \bar B_{ab}}{\sqrt{\bar n_a \sigma^2_{p_t,a} \bar n_b  \sigma^2_{p_t,b}}}.
\eea
For single bin $a$ the equivalent of Eq.~(\ref{bab}) is
\bea
\bar B_{aa} &=& \overline{(P_{t} - n \bar p_t)_a^2}  - \bar n_a \sigma_{p_t,a}^2 \rightarrow \bar B_a,
\eea
the marginal {\em variance difference} defined in Eq.~(\ref{bbar}), and
\bea \label{raa}
r_{a} &=& \frac{\bar B_a}{\bar n_a \sigma^2_{p_t,a}}
\eea
is a per-particle ratio measure analogous to $\Delta \sigma^2_n$ in Eq.~(\ref{nnmin1}). Within a limited angular acceptance $\bar n_a$ and $\sigma^2_{p_t,a}$ may be assumed constant across the bin system and $\bar n_a \sigma_{p_t,a}^2 \rightarrow \bar n \sigma_{p_t}^2$ for all bins. 

Pearson's correlation coefficient serves as a template for fluctuation measure design. One must decide in each case what is the appropriate reference for a variance or covariance depending on the context and the {\em hypothesis to be tested}. The per-particle format appearing in Eqs.~(\ref{rab}) and (\ref{raa}) is a placeholder [as opposed to per-pair measures such as Eq.~(\ref{eq6}) and Eqs.~(\ref{eq17}) and (\ref{aliceptfluctx}) below]. Fluctuation variations relative to nucleon participant number $N_{part}$ or \nn\ binary collision number $N_{bin}$ may be studied with a simple rescaling.  Each of $n$ and $P_t$ have their own mean-value and fluctuation systematics that can and should be studied first in isolation, not as ratios that may obscure underlying physical mechanisms. 

\subsection{T fluctuations and the thermodynamic analogy} \label{tfluct}

Some proposed fluctuation measures are motivated by {\em a priori} physical assumptions relating to a temperature narrative where it is assumed that the event-wise ratio $P_t / n$ may serve as a proxy for temperature $T = E / n$ in nuclear collisions by analogy with thermodynamics, with temperature fluctuations estimated by $\sigma_T^2 \sim \sigma^2_{\langle p_t \rangle}$. For central-limit conditions one expects $\sigma^2_{\langle p_t \rangle} \rightarrow  \sigma^2_{p_t} / \bar n$, providing a reference for detecting ``non-statistical'' temperature fluctuations via difference $\sigma^2_{\langle p_t \rangle} - \sigma^2_{p_t} /\bar n$. In the large-$n$ (thermodynamic) limit all fluctuations should then decrease toward zero as an apparent manifestation of thermal equilibration that is assumed to occur at some level, at least within the interval $p_t \in [0.15,2]$ GeV/c thought to exclude dijet contributions~\cite{nojets}. 

Event-wise mean \pt\ is a ratio of RVs represented by
\bea
\langle p_t \rangle &=&  \frac{P_t}{n}
\eea
for each event of an event class. Note that the symbol $\langle p_t \rangle$ was used to represent the ensemble mean in a previous LHC analysis~\cite{alicempt} (referred to  as ``inclusive'' $\langle p_t \rangle$), a quantity denoted by $\bar p_t$ in the present study. In Ref.~\cite{aliceptfluct} the event-wise mean is denoted by $M_{EbE}(p_t)$ and the event-ensemble mean is denoted  by $M(p_t)$ following symbol definitions introduced in Ref.~\cite{na49ptfluct}.

The variance of event-wise mean \pt\ is  then
\bea \label{eq6}
 \sigma^2_{\langle p_t \rangle} &=& \overline{(\langle p_t \rangle - \bar p_t)^2}
 \\ \nonumber
 &=&  \overline{\left\{\frac{(P_t -  n \bar p_t)^2}{n^2}\right\}}.
\eea
The proposed {\em per-pair} fluctuation measure $\sigma^2_{\langle p_t \rangle} - \sigma^2_{p_t} / \bar n$ is problematic for several reasons: (a) The particle multiplicity for most collisions is small, with large relative fluctuations leading to significant but unrepresented covariances as systematic biases~\cite{jeffptfluct}; (b) per-pair measures include an extra factor $1/\bar n$ compared to per-particle measures consistent with Pearson's correlation coefficient; and (c) collision mechanisms other than ``temperature'' variation  may produce \pt\ fluctuations that would be misrepresented by the variance measure in Eq.~(\ref{eq6})~\cite{ptscale,ptedep}.

\subsection{Summary of $\bf p_t$ fluctuation measures} \label{fluctsum}

When referring specifically to the event-wise mean (RV ratio) and its fluctuations I employ symbol \mpt. More generally I refer to \pt\ fluctuations. Several \pt\ fluctuation measures have been defined at the SPS and RHIC. 

The $\Phi_{p_t}$ measure~\cite{na49ptfluct} (NA49 collaboration) was motivated by a previously-observed  $n$-$P_t$ covariance in \pp\ collisions as a possible indicator of equilibration in \aa\ collisions and is defined as a difference between r.m.s.\ quantities~\cite{marek}
\bea
\Phi_{p_t} &\equiv& \sqrt{\sigma^2_{P_t|n}/ \bar n} - \sigma_{p_t}
\\ \nonumber
\left( \sqrt{\sigma^2_{P_t|n}/ \bar n} + \sigma_{p_t}\right) \Phi_{p_t} &=&  \bar B / \bar n
\\ \nonumber
&\approx& 2 \sigma_{p_t} \Phi_{p_t}.
\eea

An initial RHIC \pt\ fluctuation analysis~\cite{jeffptfluct} (STAR collaboration) defined a related r.m.s.\ measure $\Delta \sigma_{P_t|n}$ (denoted by $\Delta \sigma_{p_t:n}$ in \cite{jeffptfluct}) to permit comparisons with $ \Phi_{p_t}$ but adopted a variance-based measure system rather than r.m.s.\ quantities
\bea
\Delta \sigma^2_{P_t|n} &=& \overline{n(\langle p_t \rangle - \hat p_t)^2} - \sigma^2_{p_t}
\\ \nonumber
&=& \overline{B/n} 
\\ \nonumber
&\equiv& 2 \sigma_{p_t} \Delta \sigma_{P_t|n}
\\ \nonumber
&\approx& 2 \sigma_{p_t} \Phi_{p_t}.
\eea
The PHENIX collaboration defined a similar  r.m.s.\ measure~\cite{phenixptfluct}
\bea \label{phen}
F_{p_t} &=& \frac{\sigma_{\langle p_t \rangle} - \sigma_{p_t} / \sqrt{\bar n}}{\sigma_{p_t}/\sqrt{\bar n}}
\\ \nonumber
&\approx& \frac{ \Delta \sigma_{P_t|n}}{\sigma_{p_t}},
\eea
where $\sigma_{p_t}/\sqrt{\bar n}$ approximates a mixed-event reference.

In subsequent STAR analysis~\cite{ptscale,ptedep} variance difference $\Delta \sigma^2_{P_t|n}$ was redefined as the statistically simpler form
\bea \label{std}
\Delta \sigma^2_{P_t|n} &\equiv& \bar B/ \bar n
\eea
equivalent to $\sigma_{p_{t},a}^2 r_a$ in Eq.~(\ref{raa}). Those related per-particle measures are based on event-wise bin-sum $P_t$ fluctuations conditional on bin-sum multiplicity $n$. Denominator $\bar n$ in Eq.~(\ref{std}) serves as a placeholder for  quantities such as $N_{part}$ or $N_{bin}$ that better probe fluctuation excess in relation to conjectured collision mechanisms.

In an alternative STAR fluctuation analysis~\cite{starvol} a \mpt\ fluctuation measure was based on the assumed temperature narrative described in Sec.~\ref{tfluct} 
\bea \label{eq17}
\sigma^2_{p_t,dyn} &\equiv& \overline{  \left\{\frac{B}{n(n-1)}\right\}  }
\\ \nonumber
&\approx & \sigma^2_{\langle p_t \rangle} - \sigma^2_{p_t}/\bar n
\\ \nonumber
&\approx&  \Delta \sigma^2_{P_t|n} / (\bar n + \Delta \sigma^2_n) + \text{covariances},
\eea
which includes a dominant $1/\bar n$ trend even if the underlying correlations do not change with system size or scale. 

The main subject of the present study is the recent ALICE $\langle p_t \rangle$ fluctuation analysis~\cite{aliceptfluct} employing a measure similar to the alternative STAR measure in Eq.~(\ref{eq17})
\bea \label{aliceptfluctx}
C &\equiv& \frac{\bar B}{\overline{n(n-1)}} \approx \frac{\bar B}{\bar n^2}
\\ \nonumber
&=& \Delta \sigma^2_{P_t|n}/ (\bar n + \Delta \sigma^2_n).
\\ \nonumber
&\approx & \sigma^2_{\langle p_t \rangle} - \sigma^2_{p_t}/\bar n.
\eea
$C$ is referred to as a ``correlator'' intended to represent covariances averaged over particle pairs, but quantity $\bar B$ is actually a variance difference as demonstrated in Eq.~(\ref{bbar}), not a covariance. 
Note that $\bar n + \Delta \sigma_n^2 \rightarrow \bar n (1 + \Delta \sigma^2_n / \bar n)$ and the second term (a per-pair fluctuation measure) becomes small compared to 1 for larger multiplicities.

The measure of \mpt\ fluctuation ``strength'' actually defined in Ref.~\cite{aliceptfluct} is the r.m.s.\ quantity
\bea
\frac{\sqrt{C'}}{\bar p_t'} &\approx& \left\{ \frac{\bar B'}{\bar n'^2 \bar p_t'^2} \right\}^{1/2} \approx \left\{ \frac{\bar B}{\bar P_t^2} \right\}^{1/2},
\eea
where primes indicate statistical quantities derived from accepted particles only, not corrected to a full \pt\ acceptance extending down to zero. Some consequences of a low-\pt\ cut limiting the \pt\ acceptance are discussed in Sec.~\ref{mpttcm}, and distinctions are maintained between primed and unprimed quantities in what follows.

Ratios of statistical quantities, even mean values, may obscure information or combine systematic effects in confusing ways (e.g.\ variance difference $B$ divided by $\bar p_t$ or $\sigma^2_{p_t}$). Fluctuation systematics reflecting strong deviations from statistical references may be obscured by such ratios. Systematic trends of primary RVs representing {\em extensive} quantities (e.g.\ $n_{ch}$, $P_t$, $B$) should be considered in isolation before any ratios are introduced. 



\section{Collision models} \label{collmod}

Basic collision models include the Glauber model of \aa\ collision geometry, the two-component model of \pp\ and \aa\ hadron production near mid-rapidity and the specific TCM for hadron yields vs \aa\ centrality.

\subsection{\aa\ collision-geometry model}

Interpreting statistical trends vs nucleus-nucleus (\aa) collision centrality requires a model for the \aa\ collision geometry. The Glauber model can be used to relate certain \aa\ geometry parameters to charge multiplicity \nch\ (integrated within some angular acceptance) via the \aa\ total cross section~\cite{powerlaw}.
Glauber-model parameters include $N_{part}/2$, the number of nucleons (from one  nucleus) that participate in at least one \nn\ encounter, and $N_{bin}$, the number of \nn\ binary encounters (based on assumed \nn\ interaction cross section $\sigma_{NN}$). In relation  to the TCM (next subsection) the preferred centrality measure is mean participant-nucleon pathlength $\nu \equiv 2 N_{bin} / N_{part}$. Also relevant are estimated impact parameter $b$ and \aa\ transverse-overlap eccentricity $\epsilon$.

One should note that \pp\ collisions are not described by the eikonal approximation basic to the \aa\ Glauber model wherein  $N_{bin} \propto N_{part}^{4/3}$. For \pp\ collisions the analog to $N_{part}$ is $n_s \propto$ the number of participating low-$x$ gluons and the analog to $N_{bin}$ is $n_h \propto n_s^2 \propto$ the number of participant gluon-gluon encounters. In either case the number of participant binary collisions $\propto N_{bin}$ or $n_s^2$ predicts the nominal TCM dijet production rate~\cite{jetspec2}.

\subsection{Two-component model -- TCM} \label{tcmintro}

The \pp\ TCM was initially based on phenomenological spectrum analysis~\cite{ppprd} but has since been related to dijet production and QCD theory~\cite{fragevo,jetspec,jetspec2}. The TCM has been quite successful in describing a variety of RHIC and LHC \pp, \pa\ and \aa\ data~\cite{ppprd,hardspec,anomalous,tommpt}.
The TCM for yields, spectra and correlations is based on the assumption that hadron production near mid-rapidity proceeds via soft (projectile-nucleon dissociation) and hard (dijet production) mechanisms assumed to be linearly independent. 
The soft component is assumed to be universal, the same for all systems and collision energies. The hard component representing minimum-bias (MB) dijets follows a non-eikonal trend for \pp\ collisions and an eikonal trend for \aa\ collisions with larger A~\cite{jetspec2}. The trend for \pa\ collisions is not known a priori but may involve a smooth transition from \pp\ (\nn) to \aa~\cite{tommpt}. 

For produced quantity $X$ (e.g.\ extensive variable $n_{ch}$, $P_t$ or $B$) the two-component model (TCM) is expressed by
\bea \label{xtcm}
X &=& n_s x_s + n_h x_h(\sqrt{s})  = X_s + X_h~~\text{for \pp}
\\ \nonumber
&=& (N_{part}/2) X_s + N_{bin} X_h ~~~\text{for \aa},
\eea 
where $x_s$ and $x_h(\sqrt{s})$ in \pp\ collisions or $X_s$ and $X_h$ in \aa\ collisions are fixed quantities in a linear-superposition (LS for \pp) or Glauber linear superposition (GLS for \aa) model but may deviate from a fixed TCM reference for real collisions [e.g.\ variation of $X_h(\nu)$ with $\nu$ for more-central \aa\ collisions]. The argument of $x_h(\sqrt{s})$ indicates that the hard component of $X$ can have an energy dependence relating to the underlying scattered-parton spectrum whereas soft component $x_s$ typically does not. GLS represents eikonal linear superposition of participant \nn\ pairs in \aa\ collisions whereas LS represents non-eikonal linear superposition of participant low-$x$ gluon-gluon pairs in \pp\ collisions.

Participant number  $N_{part}$ in \aa\ collisions corresponds by hypothesis to $n_s$ in \pp\ collisions, and $N_{bin}$ corresponds to $n_h \propto n_s^2$. Thus, $\nu = 2 N_{bin} / N_{part} \propto N_{part}^{1/3}$ for \aa\ collisions corresponds to $n_h/n_s \propto n_s^2 / n_s = n_s$ for \pp\ collisions. {\em Per-participant} scaling of $X / n_s$ or $2X / N_{part}$, proportional to $n_s$ or $\nu$ respectively in the \pp\ and \aa\ systems, is interpreted as an indication of dijet production within a linear-superposition scenario.

The TCM relation between \pp\ and \aa\ collisions can be expressed by
\bea
\frac{2}{N_{part}}X &\approx &  X_{pp} + (\nu - 1) X_h' ~~~\text{for \aa},
\eea
where $X_{pp} = X_s + X_h$ for NSD or MB \pp\ $\approx$ \nn\ collisions and $X_h'$ represents a modified \nn\ hard component for secondary \nn\ scatters in \aa\ collisions~\cite{tommpt}.

The energy dependence of  the TCM is based on the empirical relation $n_s \propto \log(\sqrt{s} / \text{10 GeV})$ for the soft component (number of low-$x$ gluons near $\eta = 0$)~\cite{jetspec2}. The multiplicity hard component $n_h$ ($\approx$ dijet fragment yield) for \pp\ collisions then scales as $n_h \propto \log^2(\sqrt{s} / \text{10 GeV})$, where intercept 10 GeV is inferred from dijet systematics~\cite{ptedep,anomalous}. For $X = \bar P_t$ in \pp\ collisions soft component $\bar p_{t,s}$ is observed to be constant over a large energy interval, whereas hard component $\bar p_{t,h}$ is linearly related to the MB jet spectrum width $\propto  \log(\sqrt{s} / \text{3 GeV})$~\cite{tommpt} (and see Fig.~\ref{endep}, right).  Jet-related \pt\ angular correlations inferred from per-particle \pt\ fluctuation measure  $\Delta \sigma^2_{P_t|n} = \bar B / \bar n_{ch}$ are also observed to scale  $\propto \log(\sqrt{s} / \text{10 GeV})$~\cite{ptedep}.




\subsection{Hadron production model} \label{hadroprod}

\begin{figure}[h]
  \includegraphics[width=1.65in]{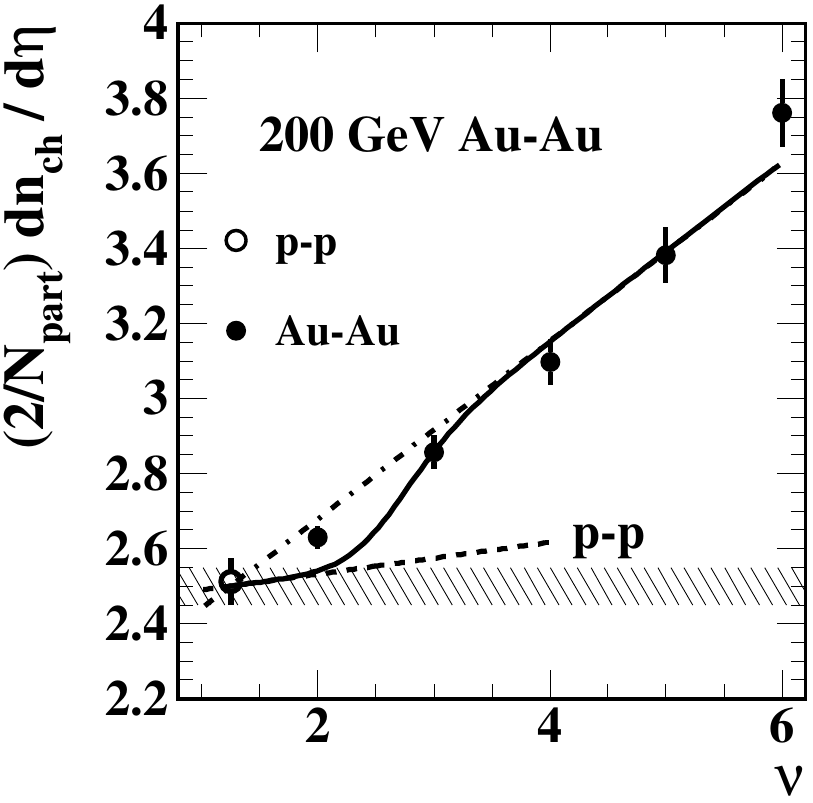}
 \put(-20,50) {\bf (a)} 
 \includegraphics[width=1.65in,height=1.63in]{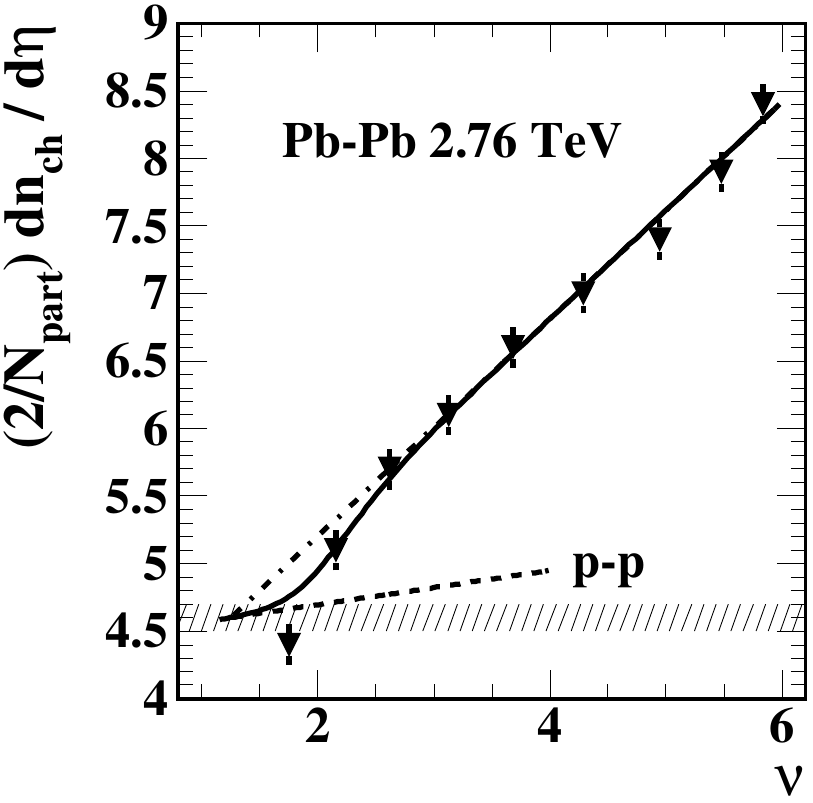}
 \put(-20,50) {\bf (b)} 
 \caption{\label{alicench}
 Left:  Per-participant hadron production measured by $(2/N_{part}) dn_{ch}/d\eta$ vs $\nu$ for 200 GeV \auau\ collisions (points) inferred from analysis of identified-hadron spectra~\cite{hardspec}. The dash-dotted line is the conventional TCM with fixed $x = 0.095$~\cite{kn}. The solid curve is described in the  text.
 Right: Hadron production vs  $\nu$ for 2.76 TeV \pbpb\ collisions from Ref.~\cite{alicench} (solid triangles). The dash-dotted line and solid curve are the TCM for 200 GeV \auau\ collisions scaled up by factors 1.84 (soft component) and $1.84^2$(hard component) reflecting soft multiplicity $n_s$ scaling as $\ln(\sqrt{s} / \text{10 GeV})$~\cite{tomphenix,jetspec2}.
  }  
 \end{figure}

Figure~\ref{alicench} (left)  shows hadron yields obtained directly from identified-hadron spectra (points~\cite{hardspec}) for   200 GeV \auau\ collisions.  The dash-dotted line shows the conventional 200 GeV \auau\ TCM with fixed $x = 0.095$~\cite{kn}. The dashed line is the GLS reference with $x = 0.006 \times 2.5 = 0.015$ predicted from \pp\ spectrum data~\cite{ppprd}. The solid curve that describes a smooth transition from one limiting case to another is defined by
  \bea \label{nchalice}
\frac{2}{N_{part}}  n_{ch} &=& n_{pp} [1 +  x(\nu) (\nu - 1.25)]
\\ \nonumber
x(\nu) &=& x_0 + x_1 \{1 + \tanh[(\nu - \nu_0)]\}/2.
  \eea
 For \auau\ data $n_{pp} \approx 2.5$, $x_0 = 0.015$, $x_1 = 0.08$, and $\nu_0 = 2.5$ represents the sharp transition from Ref.~\cite{anomalous}. 

Figure~\ref{alicench} (right) shows 2.76 TeV \pbpb\ hadron production data from Ref.~\cite{alicench} (points) compared to a corresponding TCM scaled up from the 200 GeV \auau\ trend (solid curve). For 2.76 TeV factor $1.84\approx  \ln(2760 / 10) / \ln(200/10)$ predicts the expected increase in $n_{s,NSD} \approx n_{pp} \rightarrow 4.6$, scaling with low-$x$ parton participants as described in Ref.~\cite{jetspec2}. The same factor is applied to $x(\nu) \propto n_s$. The functional form of $x(\nu)$ at 2.76 TeV is very similar to that at 200 GeV with the exception that the sharp transition (ST) in jet structure near $\nu = 3$ first reported in Ref.~\cite{anomalous} is shifted down to $\nu \approx 2$ at the higher energy, as first noted in Ref.~\cite{tomphenix}. 

Equation~(\ref{nchalice}) is used in this study to relate reported \nch\ values from Ref.~\cite{aliceptfluct} to fractional cross sections and then to Glauber parameters $N_{part}/2$, $N_{bin}$ and $\nu = 2N_{bin}/N_{part}$ according to the methods in Ref.~\cite{powerlaw}.

 \section{$\bf \bar p_t $ TCM for $\bf p$-$\bf p$ collisions} \label{mpttcm}
 
The TCM for previously-published $\bar p_t$ (ensemble-mean) data is required to process the \pt\ fluctuation data in this study. We summarize TCM results from Ref.~\cite{tommpt} that describe accurately the required \pp\ $\bar p_t$ data. The $\bar p_t$ TCM for \pbpb\ data is summarized in App.~\ref{aatcm}.

\subsection{$\bf \bar p_t$ TCM definition} \label{primes}

The  $\bar p_t$ TCM for \pp\ collisions is based on  total \pt\ denoted by $P_t$ integrated over all particles within some angular acceptance. If \nch\ is the mean total charge integrated {\em within the same acceptance} ensemble-mean $\bar p_t = \bar P_t / n_{ch}$. Just as $n_{ch} = n_s + n_h$ I assume $P_t = P_{t,s} + P_{t,h}$. The TCM for $\bar P_t(n_s)$ is then $n_s \bar p_{t,s} +  n_h\bar p_{t,h}$ and for $\bar p_t$ is
 \bea \label{pptcm}
  \bar p_{t}(n_s,\sqrt{s})
  &=& \frac{ n_s \bar p_{t,s} +  n_h \bar p_{t,h}(\sqrt{s})}{n_s + n_h} \\ \nonumber
  &=& \frac{\bar p_{t,s} + x (n_s)\bar p_{t,h}(\sqrt{s})}{1 + x(n_s)},  
\eea
where $x(n_s) = \alpha  n_s$ with $\alpha \approx 0.006$~\cite{ppprd}. $n_s$ is obtained from $n_{ch}$ by   $n_s = (1/2\alpha) [\sqrt{1+4\alpha n_{ch}} - 1]$ since $n_{ch} = n_s + \alpha n_s^2$. The two $\bar P_t$ TCM components can be inferred from \pt\ spectrum TCM model functions or from $\bar p_t$ data.
  
If the \pt\ spectrum is cut off at some small value (e.g.\ $p_{t,cut} \approx$ 0.15 GeV/c)
    \bea \label{pptcmp}
   \bar p'_{t}(n_s,\sqrt{s}) &=& \frac{n'_s \bar p'_{t,s} + n_h \bar p_{t,h}(\sqrt{s})}{n_s' + n_h}
   \\ \nonumber
   &\approx&\frac{ \bar p_{t,s} + x(n_s) \bar p_{t,h}(\sqrt{s})}{n_s'/n_s + x(n_s)},
   \eea
assuming no loss from the hard components of  \nch\ and $P_t$. We observe that the product $\bar P'_{t,s} = n'_s \bar p'_{t,s} \approx n_s \bar p_{t,s} = \bar P_{t,s}$ is (and $P_t$ fluctuations are) insensitive to a low-\pt\ cutoff because only a small fraction of integrated $P_{t,s}$ (and none of $P_{t,h}$) is involved whereas a substantial fraction of $n_{s}$ may be affected. Thus,  only the ratios $\bar p'_{t,s} \approx \bar P_{t,s} / n'_{s}$ and $n'_s / n_s$ are sensitive to a low-\pt\ cutoff.

In this study I correct affected quantities for a low-\pt\ cutoff at $p_{t,cut} \approx 0.18$ GeV/c based on Ref.~\cite{tommpt}. From reported (and corrected) $n_{ch}$ in one unit of $\eta$ I obtain $n_s$ from $n_{ch}$ as defined above. Then $n_h = n_{ch} - n_s$ and $n_s' = 0.75 n_s$ giving the sum $n_{ch}' = n_s' + h_h$. The corrected ensemble-mean \pt\ is $\bar p_t \approx (n_{ch}' / n_{ch} \bar ) \, \bar p_t'$ given $\bar P_t \approx \bar P_t'$.

\subsection{TCM description of  $\bf \bar p_t$ data}

\begin{figure}[h]
 \includegraphics[width=1.65in,height=1.6in]{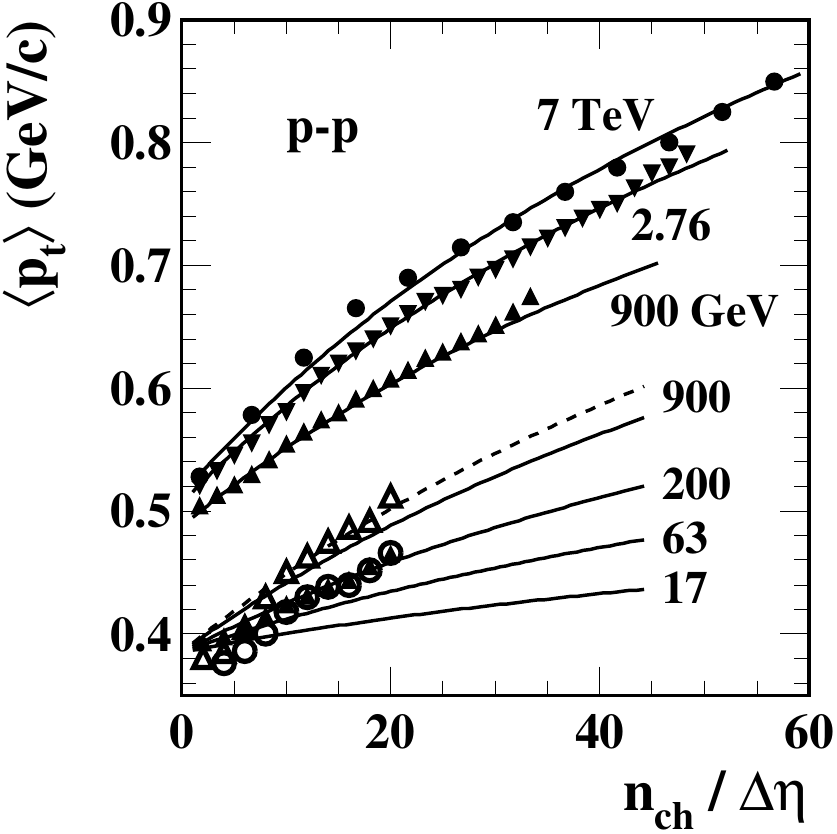}
 \includegraphics[width=1.65in,height=1.6in]{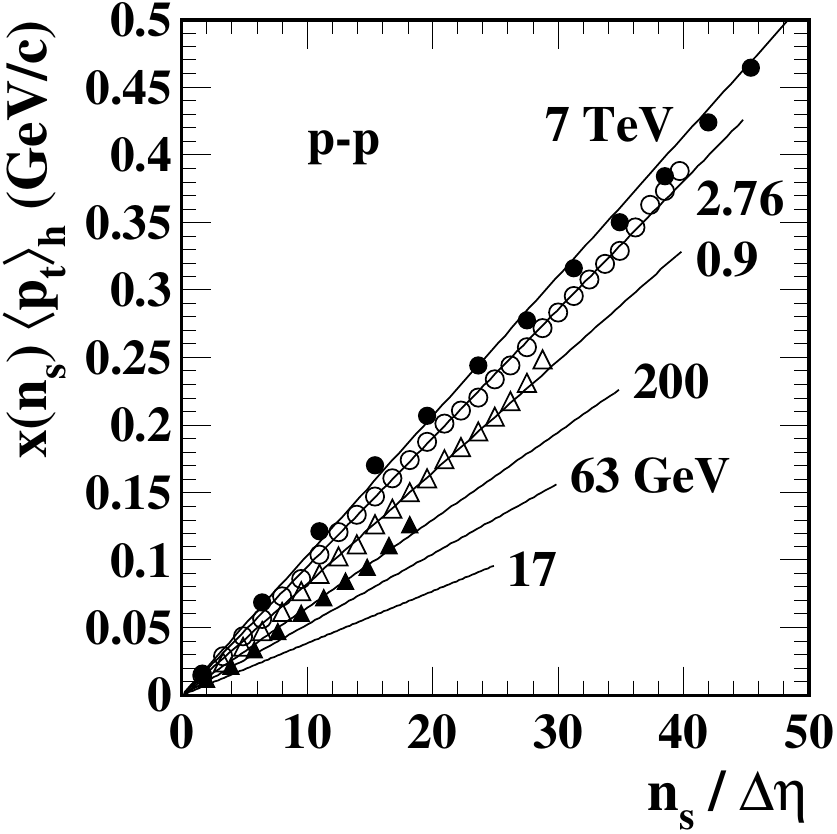}
\caption{\label{mptdat}
Left: $\bar p_t$ vs \nch\ for several collision energies  (represented as $\langle p_t \rangle$ in Ref.~\cite{tommpt}). The upper group of points is $\bar p_t'$ from Ref.~\cite{alicempt} biased by a low-\pt\ acceptance cut near 0.15 GeV/c. 
Right: LHC $\bar p_t'$ data from the left panel multiplied by factor ${n'_{ch}} / n_s$ that removes the bias from the low-\pt\ cut and the jet contribution to the denominator of $\bar p_t'$. The universal soft component $\bar p_{t,s}$ is then subtracted according to Eq.~(\ref{mptrat}) isolating the $\bar P_t$ hard component as $x(n_s) \bar p_{t,h}(\sqrt{s}) = \bar P_{t,h} / n_s$.
 }  
\end{figure}

Figure~\ref{mptdat} (left) shows LHC $\bar p_t'$ data from \pp\ collisions at 0.9, 2.76 and 7 TeV (upper points)~\cite{alicempt}. The LHC particle data were obtained with a nominal $p_{t,cut} = 0.15$ GeV/c, but multiplicity \nch\ was extrapolated to zero \pt. Also included are reference $\bar p_t$ data from UA1 (open triangles, open circles~\cite{ua1mpt}) and STAR (solid points~\cite{ppprd}) obtained by model fits to spectra.   The curves are defined by Eqs.~(\ref{pptcm}) or (\ref{pptcmp}) with parameters from Ref.~\cite{tommpt}.


Figure~\ref{mptdat} (right) shows the quantity
\bea \label{mptrat}
\frac{n'_{ch}}{n_s} \bar p_t'(\sqrt{s}) -  \bar p_{t,s} &\approx & x(n_s)  \bar p_{t,h}(\sqrt{s})
\eea
 where the expression on the right follows from Eq.~(\ref{pptcmp}), $\bar p_{t,s} = 0.385$ GeV/c is assumed for all energies  and $x(n_s) = \alpha\, n_s / \Delta \eta$ with $\alpha = 0.0055$ for $\Delta \eta = 1.6$. The first term of  $n'_{ch}/n_s = n'_s/n_s + x(n_s)$ is adjusted so that the various data sets have a common intercept at 0. The values are consistent with an {\em effective} $p_{t,cut} \approx 0.18$ GeV/c. 

 
 \subsection{$\bf \bar p_t$ energy dependence and relation to MB jets}
 
\begin{figure}[h]
     \includegraphics[width=1.65in,height=1.6in]{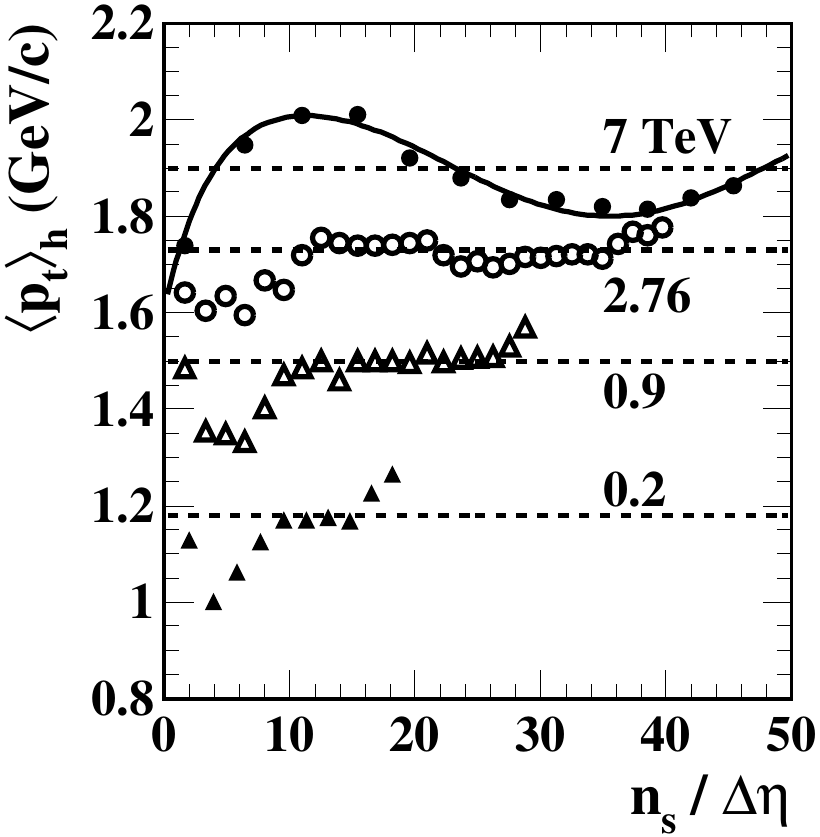}
     \includegraphics[width=1.65in,height=1.6in]{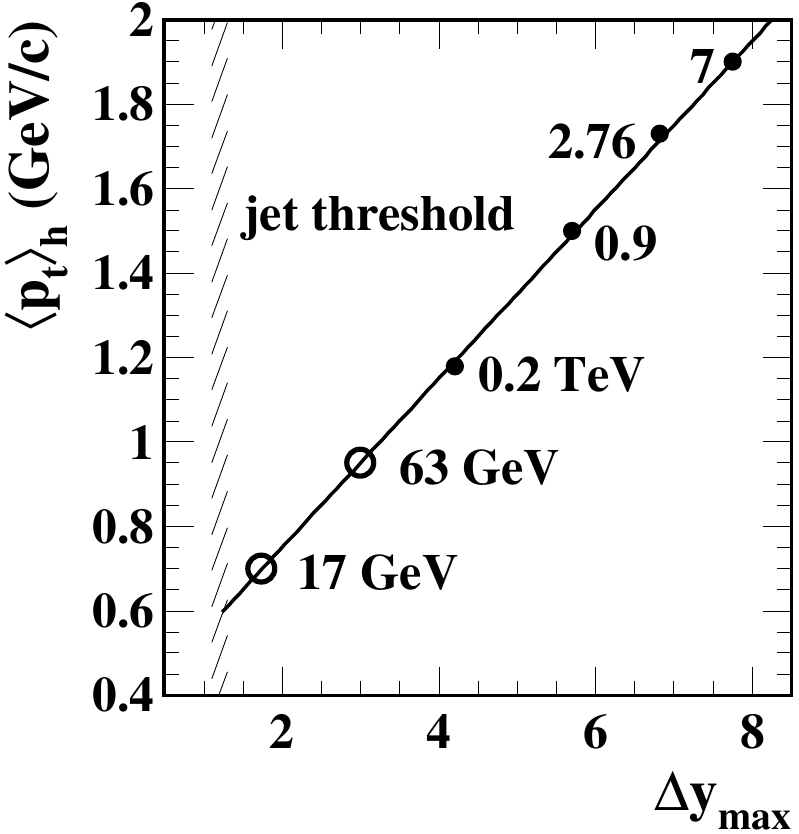}
   \caption{\label{endep}
   Left: Hard components $\bar p_{t,h}$ (represented here as $\langle p_t \rangle_h$ from Ref.~\cite{tommpt}) isolated according to Eq.~(\ref{mpthc}). Most data fall within narrow horizontal bands reflecting point-to-point data variation consistent with the \pp\ TCM. The solid curve is a parametrization of 7 TeV data used in the present study.
    Right: $\bar p_{t,h}$ mean values from the left panel plotted vs parameter $\Delta y_{max} = \ln(\sqrt{s}/\text{3 GeV})$ describing variation of minimum-bias jet-spectrum widths with \pp\ collision energy~\cite{jetspec2}.
      }  
    \end{figure}

 Figure~\ref{endep} (left) shows $\bar p_t$ data in the form
 \bea \label{mpthc}
\frac{1}{x(n_s)} \left(\frac{n'_{ch}}{n_s} \bar p_t'(\sqrt{s}) - \bar p_{t,s}\right) &=& \bar p_{t,h}(\sqrt{s})
 \eea
 for four energies, where $ \bar p_{t,s}$ has fixed value 0.385 GeV/c for all energies. Most of the $\bar p_{t,h}$ values fall in narrow horizontal bands, but the significant downturn for smaller multiplicities is a real feature of the spectrum hard component first observed for spectra from 200 GeV  \pp\ collisions with smaller \nch~\cite{ppprd}. The solid curve is a 7 TeV parametrization used in the present study.
 
 Figure~\ref{endep} (right) shows $\bar p_{t,h}(\sqrt{s})$ mean values from the left panel (solid points) vs  $\Delta y_{max} = \ln(\sqrt{s} / \text{3 GeV})$ from Ref.~\cite{jetspec2} where it was demonstrated that jet spectrum widths scale with \pp\ collision energy as $\Delta y_{max}$. The right panel reveals that $\bar p_{t,h}$ is linearly related to the MB jet spectrum width. That trend is consistent with the results of Ref.~\cite{fragevo} where it was demonstrated that the spectrum hard component that determines $\bar p_{t,h}$ is predicted by folding an ensemble of parton fragmentation functions with a MB jet spectrum. The hadron spectrum hard-component width should then scale linearly with the jet spectrum width, and $\bar p_{t,h}$ should have the linear relation to $\Delta y_{max}$ demonstrated above.  For \pp\ collisions the $\bar p_t$ vs \nch\ systematics  compel a jet interpretation for the TCM hard component. The $\bar p_t$ soft component remains consistent with longitudinal-projectile (nucleon) fragmentation independent of collision system or energy. Open symbols are predictions for lower energies. 

    
 

\section{LHC $\bf p$-$\bf p$ $\bf p_t$ fluctuations} \label{alicempt}

In Ref.~\cite{aliceptfluct} \pp\ \mpt\ fluctuation data are reported for $\sqrt{s} = 0.9$, 2.76 and 7 TeV.  Plotted as $\sqrt{C'} / \bar p_t'$ vs $n_{ch}$ the data for different energies are not distinguishable (Fig.~1 of Ref.~\cite{aliceptfluct}). We first focus on the 7 TeV data and later estimate the energy dependence of variance difference $B$. 

\subsection{$\bf p$-$\bf p$ $\bf p_t$-fluctuation multiplicity dependence} \label{ppmult}
 
\begin{figure}[h]
   \includegraphics[width=1.65in,height=1.6in]{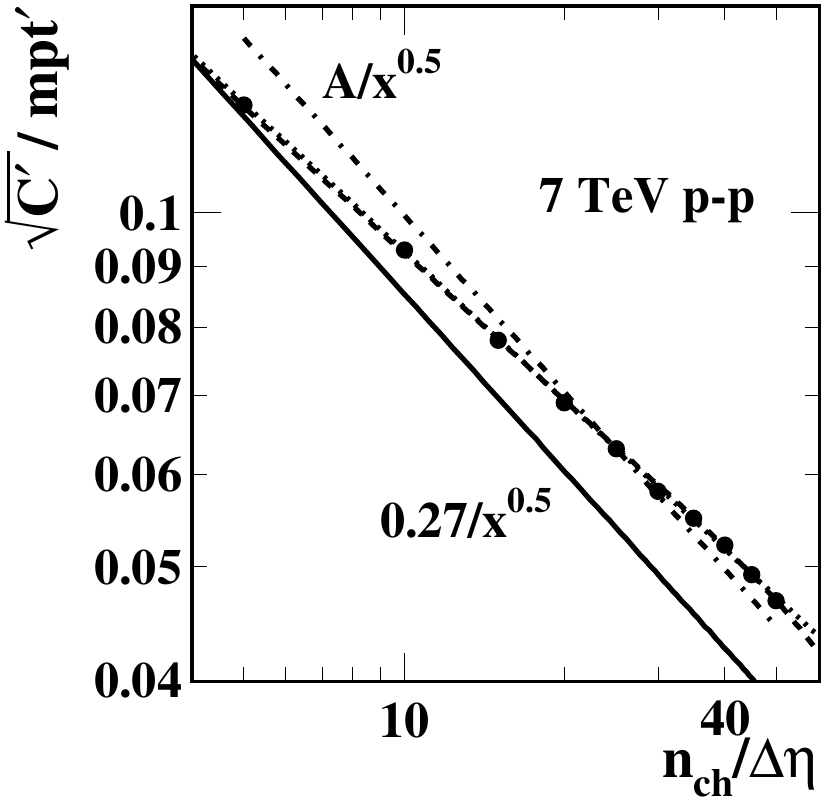}
   \includegraphics[width=1.65in,height=1.63in]{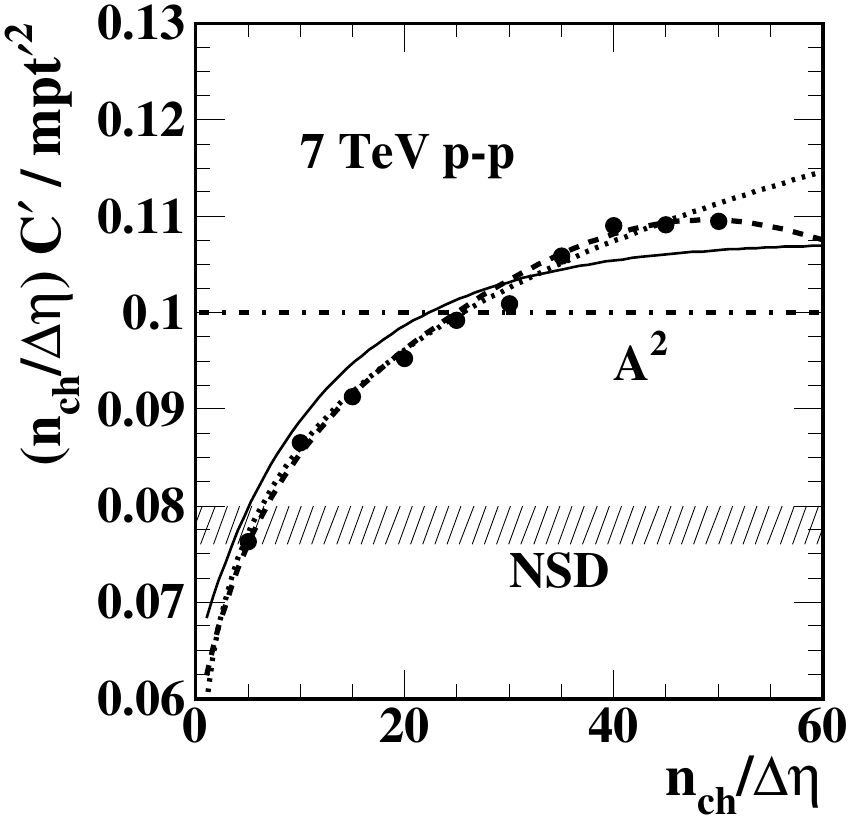}
 \caption{\label{pp1}
 Left: Representation of \mpt\ fluctuation data from Ref.~\cite{aliceptfluct} Fig.~1 (points) with a measure that exhibits similar trends for three LHC energies. The curves are described in the text. $\text{mpt}'$ stands for uncorrected $\bar p_t'$ biased by a low-\pt\ acceptance cut.
  Right:  Data from the left panel transformed according to the axis labels but employing corrected charge density $n_{ch} / \Delta \eta$. The curves are described in the text.
    } 
  \end{figure}

Figure~\ref{pp1} (left) shows $\sqrt{C'} / \bar p_t'$ data (points) vs corrected $n_{ch}$ from Ref.~\cite{aliceptfluct} (Fig.\ 4 left) plotted in a log-log format. A ``power law'' trend $\propto 1/\sqrt{n_{ch}}$ noted in Ref.~\cite{aliceptfluct} is also shown (solid line). In Ref.~\cite{aliceptfluct} \pp\ data were fitted with a power-law model function $\propto 1/n_{ch}^b$ with $b \approx 0.4$. The dotted curve is $0.244/n_{ch}^{0.42}$. The dash-dotted curve represents a fit to HIJING data reported in Ref.~\cite{aliceptfluct}. The dashed curve is explained below. Among several plotting formats appearing in the present study this format is comparatively insensitive to information in the data.

Figure~\ref{pp1} (right) shows $n_{ch}  C' / \bar {p}_t'^2 \approx B/ n_{ch}\bar p_t^2$. The data (points) increase monotonically with multiplicity over a 30\% interval. The dotted curve is the power-law model in the left panel suitably transformed to obtain a trend $\propto n_{ch}^{0.16}$. The dash-dotted line is $A^2 = 0.315^2$ from the left panel. The hatched band represents $(B_{pp} / n_{ch})/ \bar p_t^2 \approx 0.0145 / 0.43^2 = 0.078$ with the NSD \pp\ value $B_{pp}$ defined in Sec.~\ref{ppedepp}. The dashed curve is explained below. This format is more sensitive to information in the data but does not reveal the underlying production mechanisms.

The $\bar p_t$ TCM summarized in Sec.~\ref{mpttcm} demonstrated that $\bar p_t$ vs $n_{ch}$ systematics (rapidly increasing for \pp\ collisions) are mainly controlled by dijet production~\cite{tommpt}. The denominator $\bar P_t^2$ in $C'/ \bar p_t'^2 \approx B / \bar P_t^2$ may then obscure the mechanisms driving variance difference $B$ in the numerator. To better isolate underlying \pt\ fluctuation mechanisms one should remove the extraneous factors $\bar p_t'^2$ and $n_{ch}'^2$ from the ALICE measure to obtain variance difference $B$ as an extensive \pt\ fluctuation measure.

\begin{figure}[h]
    \includegraphics[width=1.65in,height=1.6in]{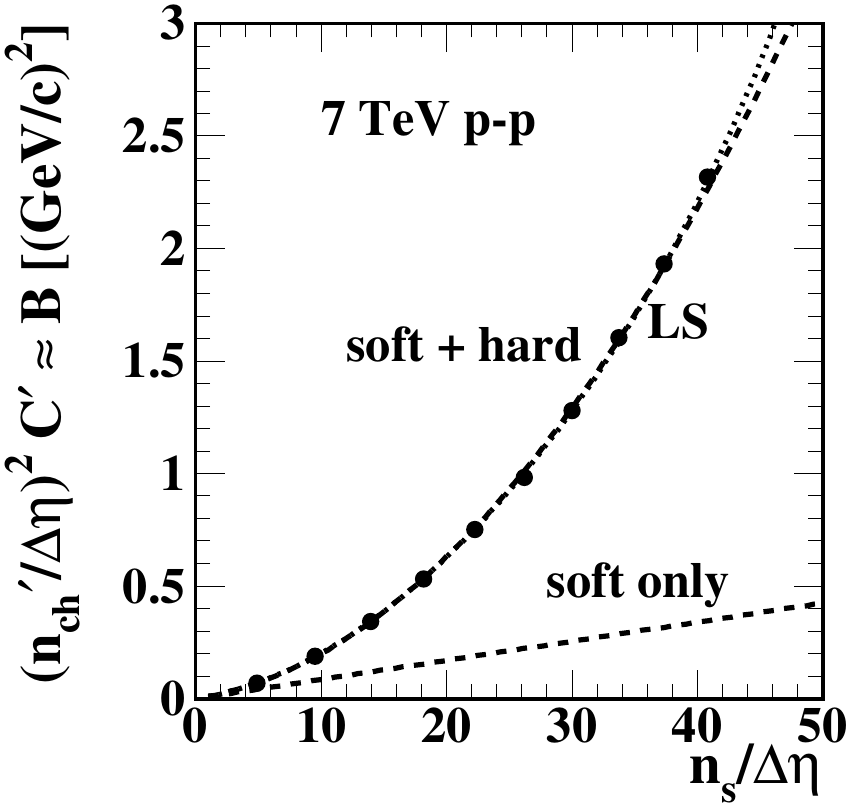}
    \includegraphics[width=1.65in,height=1.6in]{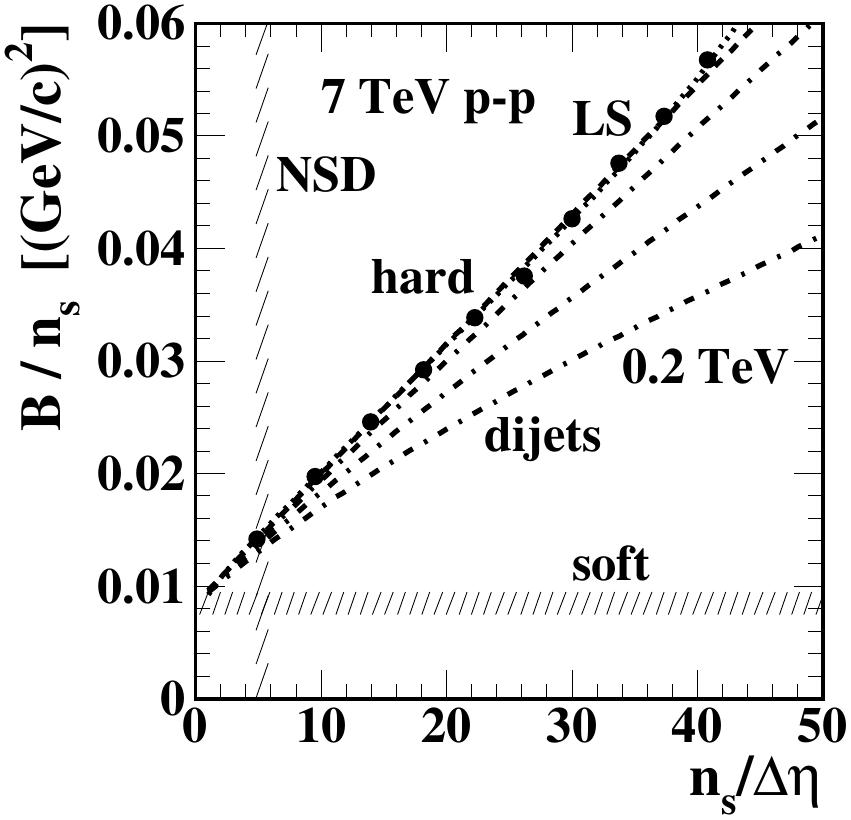}
  \caption{\label{pp2}
  Left: $P_t|n$ variance difference $B$ derived from the data in Fig.~\ref{pp1} (left) according to the axis labels. Factor $n_{ch}'^2$ biased by a low-\pt\ acceptance cut cancels the same bias in $C'$.
   Right: Ratio $B/n_s$ data (points) exhibiting the LS scaling (dashed line) expected for dijets.  In both panels the dashed curve (line) represents the LS trend (participant-gluon-pair linear superposition) in the \pp\ TCM.
     } 
   \end{figure}

Figure~\ref{pp2} (left) shows $n_{ch}'^2 C' \approx B$ vs soft multiplicity $n_s$ (points) with $n_s$ inferred from $n_{ch}$ as described above. The transformation from $\sqrt{C'}/\bar p_t'$ data to $C'$ is based on the $\bar p_t'$ TCM described in Sec.~\ref{mpttcm}, including the 7 TeV parametrization of $\bar p_{t,h}$ (solid curve) in Fig.~\ref{endep} (left). The dashed curve LS through the data is explained below. The dotted curve is the ``power-law'' expression from Fig.~\ref{pp1} (left, dotted curve) suitably transformed.

A striking feature of the $B$ trend is the large values for higher multiplicities. Both dijet production and the \pt-fluctuations hard component scale as $n_s^2$, increasing 70-fold in the $n_s$ interval from 5.3 (NSD) to 45~\cite{ppprd}. At the upper $n_s$ limit of the $B$ data the hard-component multiplicity is about 25\% of the soft component, and several jets (with mean energy $\approx 5$ GeV) may appear {\em per collision} within $\Delta \eta = 1$~\cite{jetspec}.  For $n_s = 45$ $\bar P_{t,h} \approx 20$ GeV/c (vs $\bar P_{t,s} \approx 17$ GeV/c) and $B \approx 2.5$ (GeV/c)$^2$. 
The large $B$ values are then consistent with \pp\ collision dynamics dominated by dijet production.\\

Figure~\ref{pp2} (right) shows the same data (points) in the form of ratio $B/ n_s $ vs $ n_s$.  The dashed line is $B/n_s$ derived from a TCM for $B$ similar to the $\bar p_t$ results of Ref.~\cite{tommpt}
\bea \label{bns}
B &=& n_s b_s + n_h b_h = B_s + B_h
\eea
as in Eq.~(\ref{xtcm}),
with $b_s = 0.0085$  (GeV/c)$^2$ and $b_h \approx 0.21$  (GeV/c)$^2$ assuming $n_h = \alpha n_s^2$ with $\alpha = 0.0055$~\cite{ppprd}. That relation represents the TCM reference for \pt\ fluctuations corresponding to $n_{ch}$ systematics in Ref.~\cite{ppprd}: linear superposition of participant low-$x$ gluon-gluon collisions within \pp\ collisions. For $\bar P_t$ expressed in the same TCM format [numerator of Eq.~(\ref{pptcm})] the coefficients are $\bar p_{t,s} = 0.385$ GeV/c and $p_{t,h} \approx 1.9$ GeV/c at 7 TeV.
The dashed curves in previous panels are Eq.~(\ref{bns}) back transformed according to the various measure definitions and using the $n_{ch}'$ and $\bar p_t'$ TCM from Sec.~\ref{mpttcm}. The solid curve in Fig.~\ref{pp1} (right) is the same back transformation employing a constant value $\bar p_{t,h} = 1.9$ GeV/c rather than  the 7 TeV $\bar p_{t,h}$ parametrization (solid curve) in Fig.~\ref{endep} (left).

As with the spectrum analysis in Refs.~\cite{ppprd,hardspec} soft and hard fluctuation components {\em plotted in an appropriate format} can be isolated via the trend vs $n_{ch}$ or $n_s$. The \pp\ soft-component estimate in the present case is the hatched band in the right panel. 
The linear trend of $B / n_s$ is additional evidence for dijet production in \pp\ collisions: soft $n_s$ represents $N_{part}$ for low-$x$ participant gluons within protons (instead of participant nucleons within nuclei), and $n_s^2$ represents $N_{bin}$ for dijet production from gluon-gluon binary collisions~\cite{tommpt}. 
As inferred from increase of ensemble mean $\bar p_t$ with $n_{ch}$ in Ref.~\cite{tommpt} the {\em dominant mechanism} for variation of \pt\ fluctuations in 7 TeV \pp\ collisions is apparently MB dijet production.


\subsection{$\bf p$-$\bf p$ $\bf p_t$-fluctuations energy dependence} \label{ppedepp}

According to Ref.~\cite{aliceptfluct} quantity $\sqrt{C} / \bar p_t$ (and therefore $C/\bar p_t^2$) vs \nch\ exhibits no significant collision-energy dependence for \pp\ over a substantial  energy interval, implying that $B / n_{ch}^2 \bar p_t^2$ is also nearly invariant for those conditions or $B \sim \bar p_t^2$ (for given $n_{ch}$) in terms of energy dependence. But the energy dependence of $\bar p_t$ is well known from Refs.~\cite{alicempt,tommpt} and the energy dependence of fluctuation measure $B$ can be estimated accordingly.

Figure~\ref{pp2} (right) includes $B/n_s$ trends (dash-dotted curves) rescaled from 7 TeV by $\bar p_t^2(n_s,\sqrt{s})$ ratios to other energies (2.76, 0.9 and 0.2 TeV). The substantial energy dependence is evident especially for larger multiplicities. The figure can be compared with Fig.~\ref{mptdat} (right) showing the $\bar P_{t,h} / n_s$ energy dependence. That a \pt\ variance measure might have an energy dependence similar to that for $\bar p_t^2$ is understandable if a single underlying dijet mechanism is common to the two cases~\cite{tommpt}. 

For the analysis of \pbpb\ data below it is useful to determine the soft and hard components of $B_{pp}$ ($B$ evaluated for NSD \pp\ collisions) for several energies by estimating the  terms in Eq.~(\ref{bns}). 
The result for 7 TeV NSD \pp\ collisions with $n_{ch} / \Delta \eta \approx 5.3$ is $B_{pp} \equiv B_s + B_h = 0.045 + 0.032 = 0.077$ (GeV/c)$^2$, where $B_h$ represents the result for in-vacuum \pp\ (first-hit or primary \nn) collisions as opposed to $B_h'$ for an average of secondary \nn\ collisions within \aa\ collisions~\cite{tommpt}. The per-particle ratio is $B_{pp} / n_s \approx 0.0145$ (GeV/c)$^2$. 
The result for 2.76 TeV \pp\ collisions with $n_{ch} / \Delta \eta \approx 4.6$ is $B_{pp} \approx 0.039 + 0.024 = 0.063$ (GeV/c)$^2$ with $B_{pp} / n_s \approx 0.0135$ (GeV/c)$^2$. The result for 200 GeV \pp\ collisions with $n_{ch} / \Delta \eta \approx 2.5$ is  $B_{pp} = 0.021 + 0.007 = 0.028$ (GeV/c)$^2$ with $B_{pp} / n_s = 0.0112$ (GeV/c)$^2$. The $B_h$ estimates all assume $b_h = 0.21$ (GeV/c)$^2$ as for 7 TeV.


\section{LHC $\bf Pb$-$\bf Pb$ $\bf p_t$ fluctuations} \label{pbpbfluct}

We next consider 2.76 TeV \pbpb\ data from Fig.\ 8 of Ref.~\cite{aliceptfluct}. Those $\sqrt{C'}/\bar p_t'$ data have been divided by ``power law'' model function $A (n_{ch}/\Delta \eta)^{-0.5}$ fitted to corresponding HIJING data to obtain the constant $A \approx 0.315$. Primes denote uncorrected quantities derived from particles falling within a restricted \pt\ acceptance.


\subsection{2.76 TeV Pb-Pb vs 200 GeV Au-Au}

\begin{figure}[h]
   \includegraphics[width=1.65in,height=1.6in]{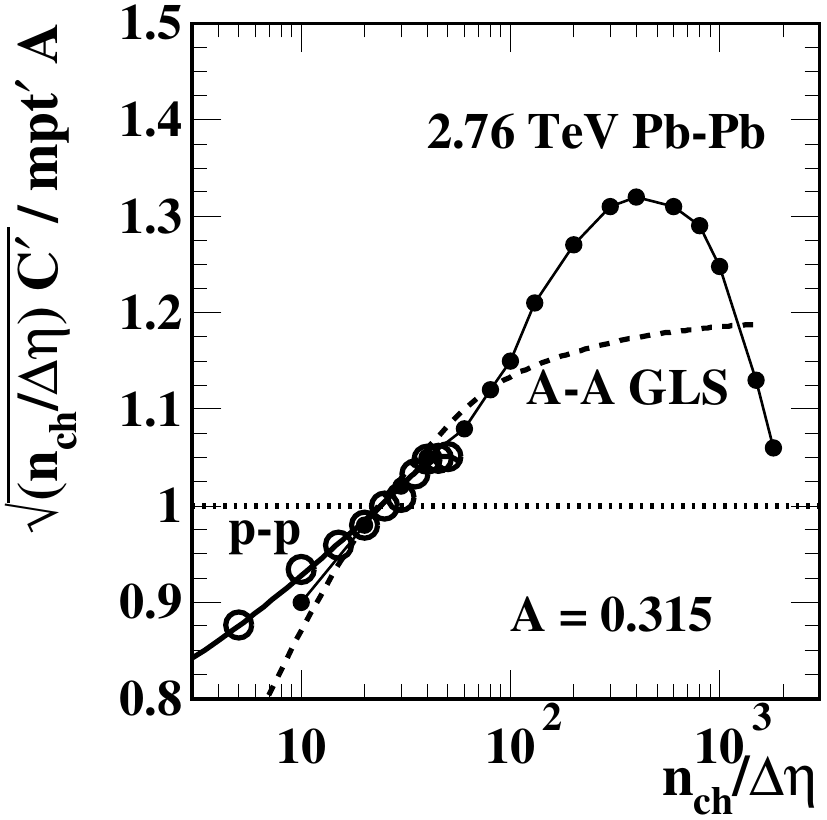}
   \includegraphics[width=1.65in,height=1.6in]{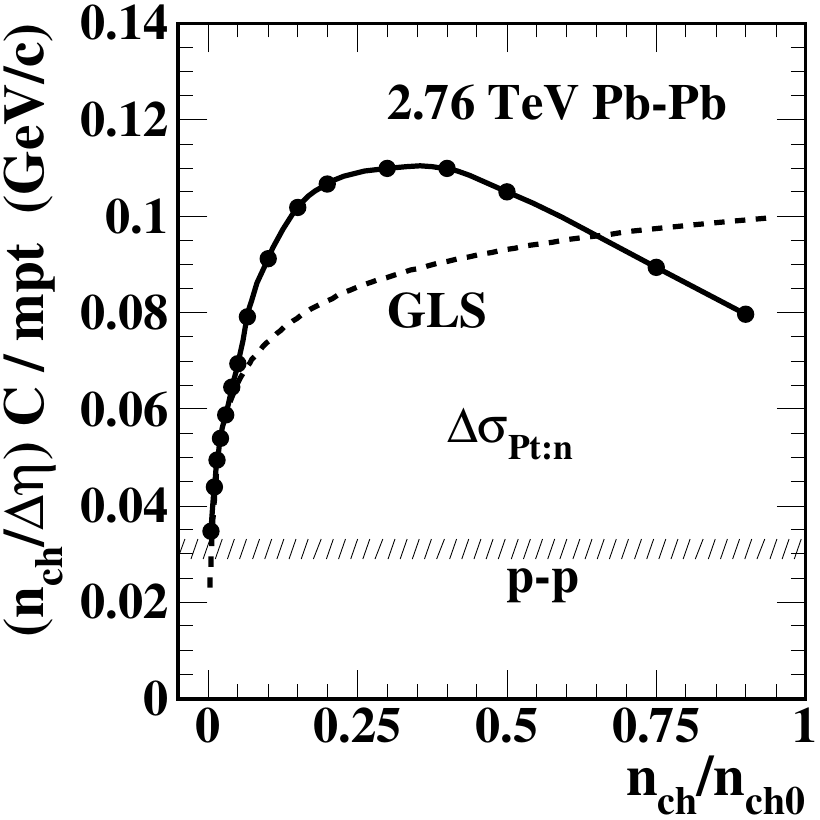}
 \caption{\label{pbpb1}
 Left: Representation of 2.76 TeV  \pbpb\ data from Fig.~8 (solid points) and 7 TeV  \pp\ data from Fig.~1 (open circles) of Ref.~\cite{aliceptfluct} normalized to $A/\sqrt{n_{ch}}$ from a fit to HIJING data.
  Right: Transformation of the \pbpb\ data from the left panel according to the axis labels with $A = 0.315$. The result is equivalent to $\Delta \sigma_{P_t|n}$ from Ref.~\cite{jeffptfluct}. $n_{ch0}$ is an estimate of the charge multiplicity corresponding to $b = 0$ central \pbpb\ collisions. The curves in both panels are described in the text.
    } 
  \end{figure}

\begin{figure}[h]
\includegraphics[width=1.65in]{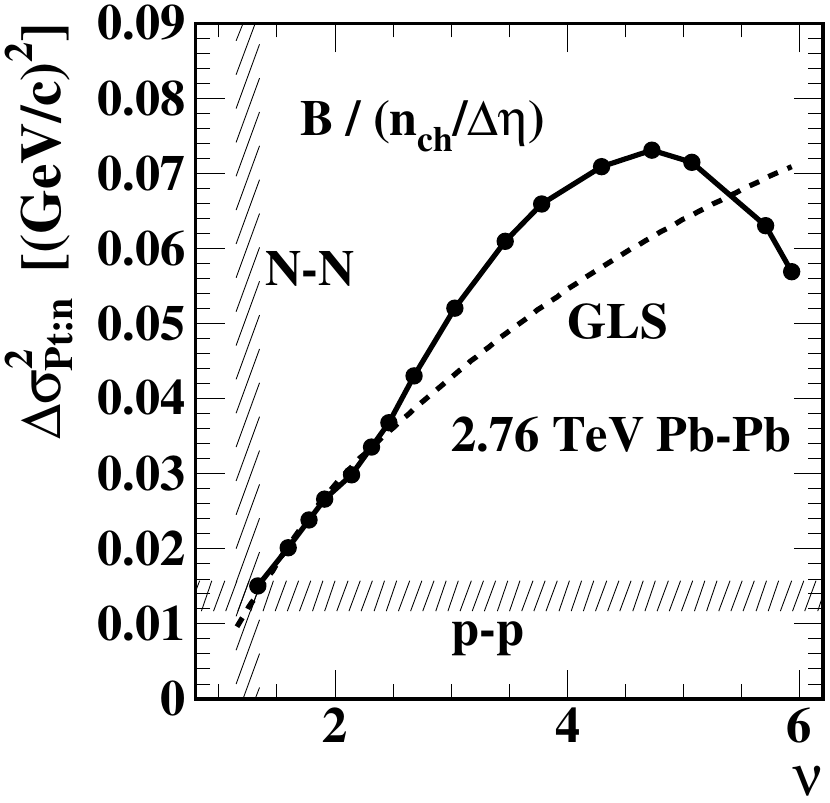}
\includegraphics[width=1.65in]{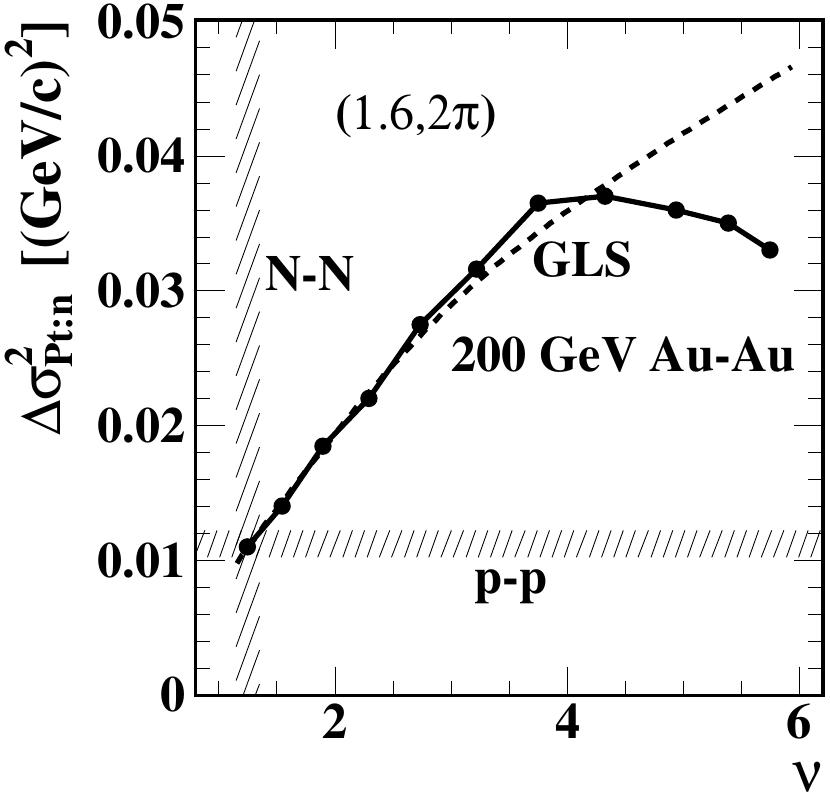}
\caption{\label{yyy}
Left: $n_{ch}C \approx B / n_{ch} = \Delta \sigma^2_{P_t|n}$ (points) transformed from the 2.76 TeV \pbpb\ data in Fig.~\ref{pbpb1} (right).
Right: Equivalent data for 200 GeV \auau\ collisions obtained from a study of \pt\ fluctuation scale dependence in Ref.~\cite{ptscale}.
} 
\end{figure}

Figure~\ref{pbpb1} (left) shows 2.76 TeV \pp\ (open circles) and \pbpb\ (solid dots) data in the form $\sqrt{(dn_{ch}/d\eta)\, C'}/\bar p_t' A$ plotted vs $n_{ch} / \Delta \eta$ representing Fig.~8 of Ref.~\cite{aliceptfluct}. While  there is a rough correspondence in the magnitude of data from the \pp\ and \pbpb\ collision systems there is a strong quantitative disagreement in the trends, just as observed for the $\bar p_t$ vs $n_{ch}$ trends in Ref.~\cite{alicempt}. 
The intersection of \pp\ and \pbpb\ trends near $n_{ch} / \Delta \eta = 25$ is misleading. Where a single \pp\ collision in that interval may produce one dijet a single \aa\ collision includes many \nn\ collisions each with the same small probability ($\ll 1$) of producing a dijet, representing very different physical contexts. Whereas $B / n_s \propto n_s$ in Fig.~\ref{pp2} (right)  $B / n_{ch} \propto n_{ch}^{1/3}$ in Fig.~\ref{yyy} (left). Interpretation is also hindered due to the r.m.s.\ nature of  the preferred ratio measure $\sqrt{C}/\bar p_t$. Below I transform the \pbpb\ data to alternative formats.

Figure~\ref{pbpb1} (right) shows the \pbpb\ data in the left panel replotted as $(n_{ch}/\Delta \eta)\, C/\bar p_t \approx \Delta \sigma_{P_t|n}$ vs $n_{ch}/n_{ch0}$ for comparison with Fig.\ 2 of Ref.~\cite{jeffptfluct} (STAR) ($n_{ch0}$ is an estimate of the multiplicity corresponding to $b = 0$ central collisions, as in Ref.~\cite{jeffptfluct}). The published data have been corrected via $n_{ch}' \rightarrow n_{ch}$ and $\bar p_t' \rightarrow \bar p_t$ according to the expressions in Sec.~\ref{primes}. The required $\bar p_t'(n_{ch})$ values were obtained from the parametrization of 2.76 TeV \pbpb\ $\bar p_t$ data introduced in Ref.~\cite{tommpt} and summarized in App.~\ref{aatcm}. The value for \pp\ collisions (hatched band) is obtained from the \pp\ NSD value $B_{pp} / n_s$ from Sec.~\ref{ppmult} and the NSD $\bar p_t$ value from Ref.~\cite{tommpt} -- $(B_{pp} / n_s) / \bar p_t = 0.0135 / 0.42 = 0.032$ GeV/c. The GLS reference (dashed curve) is explained below. There is good agreement with Ref.~\cite{jeffptfluct} 130 GeV data in general shape. The higher-energy data exhibit larger fluctuation amplitudes ($\approx $ \mbox{2-fold} increase) as  expected for a dijet mechanism. As with plots on $N_{part}$ this plot on $n_{ch}$ compresses more-peripheral data into a small interval and de-emphasizes the important GLS scaling at the left.


Reference~\cite{phenixptfluct} (PHENIX collaboration) reported \pt\ fluctuations measured by $F_{p_t} \approx \Delta \sigma_{P_t|n} / \sigma_{p_t}$ defined in Eq.~(\ref{phen}).
The centrality trend of $F_{p_t}$ on $N_{part}$ in Fig.~2 of Ref.~\cite{phenixptfluct} corresponds well with the centrality trend of $\Delta \sigma_{P_t|n}$ on $n_{ch} / n_{ch0}$ in Ref.~\cite{jeffptfluct} and to the LHC fluctuation data in the format of Fig.~\ref{pbpb1} (right). The maximum value $\sigma_{p_t} F_{p_t} \approx 0.3 \times 0.035 \approx 0.01$ GeV/c appears to conflict with  $\Delta \sigma_{P_t|n} \approx 0.05$ GeV/c reported in Ref.~\cite{jeffptfluct}. 
But total variance scales approximately with angular acceptance (e.g.\ Fig.~\ref{fluctinv} left), which for STAR is $2 \times 2\pi = 4\pi$ whereas for PHENIX (in that study) it is $0.7 \times \pi = 0.7 \pi$. The acceptance ratio 5.7 thus accounts reasonably well for the STAR/PHENIX fluctuation data ratio 5.

Reference~\cite{phenixptfluct} also presents the effect of varying the acceptance upper limit $p_\text{T}^\text{max}$  (equivalent to a running integral) in its Fig.~3, where the maximum rate of increase  occurs just above 1 GeV/c at the mode of the spectrum hard component reported in Refs.~\cite{ppprd,hardspec} and consistent with MB dijets as the principal source of \pt\ fluctuations.

Figure~\ref{yyy} (left) shows the \pbpb\ data from Fig.~\ref{pbpb1} (left) converted to per-particle variance difference $ (n_{ch}/\Delta \eta) C  \approx \Delta \sigma^2_{P_t|n} \equiv B / (n_{ch} / \Delta \eta)$ according to Eq.~(\ref{aliceptfluctx}) and plotted vs mean participant pathlength $\nu$. The general trend is strong increase with centrality. The dashed curve is explained below. When data are plotted on mean participant pathlength $\nu$ the GLS data trend is apparent for more-peripheral data. The \pbpb\ data follow the GLS reference up to $\nu = 2.5$, suggesting transparent \pbpb\ collisions within that interval~\cite{anomalous}. The 2.76 TeV \pp\ reference value is $B_{pp} / n_s \approx 0.0135$ (GeV/c)$^2$.

Figure~\ref{yyy} (right) shows equivalent data for 200 GeV \auau\ collisions from Ref.~\cite{ptscale} reporting a study of fluctuation scale (bin-size) dependence of \pt\ fluctuations (see Sec.~\ref{scale}). Those data correspond to bin size $(\Delta \eta,\Delta \phi) = (1.6,2\pi)$ matching the ALICE detector acceptance. 
The general variation with centrality is similar, but values for the higher collision energy are substantially larger as expected for a dijet production mechanism. The \auau\ data appear to follow a GLS reference (dashed curve) up to at least $\nu \approx 3$. The curvature of the GLS trend at the higher collision energy (left) is greater than that at the lower energy because of increased dijet production and a larger hard-component contribution to $n_{ch}$ in the denominator of $\Delta \sigma^2_{P_t|n} = B / n_{ch} $.
The 200 GeV \pp\ reference value is $B_{pp} / n_s \approx 0.0112$ (GeV/c)$^2$ (derived from 7 TeV \pp\ data in Sec.~\ref{ppedepp}).
We now introduce an additional factor $2n_{ch} / N_{part}$ to obtain $(2 / N_{part}) B$, the variance difference per participant-nucleon pair.

\begin{figure}[h]
\includegraphics[width=1.65in]{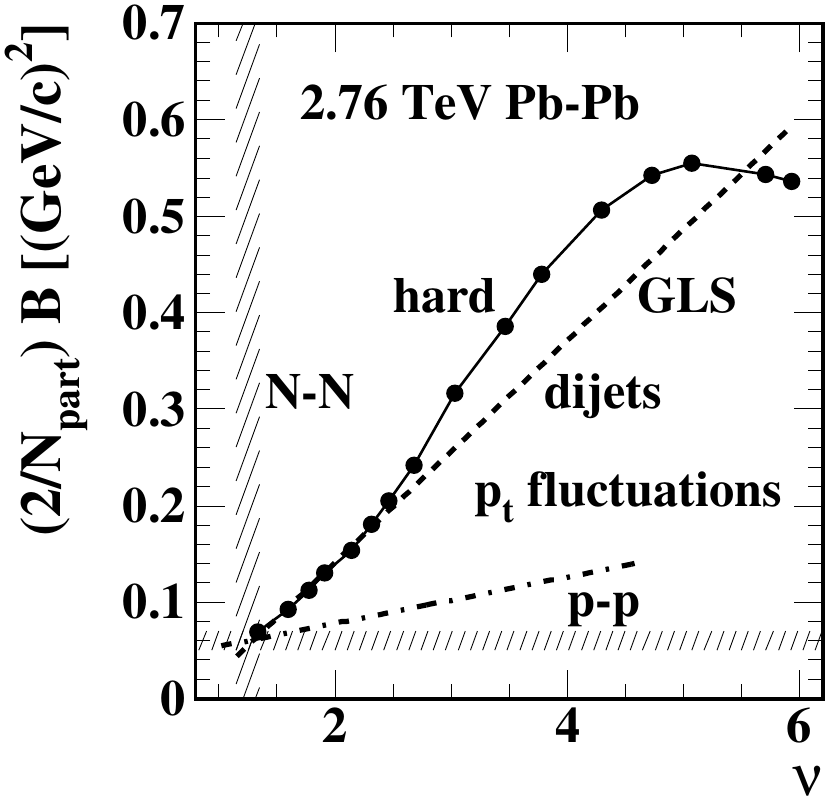}
\includegraphics[width=1.65in]{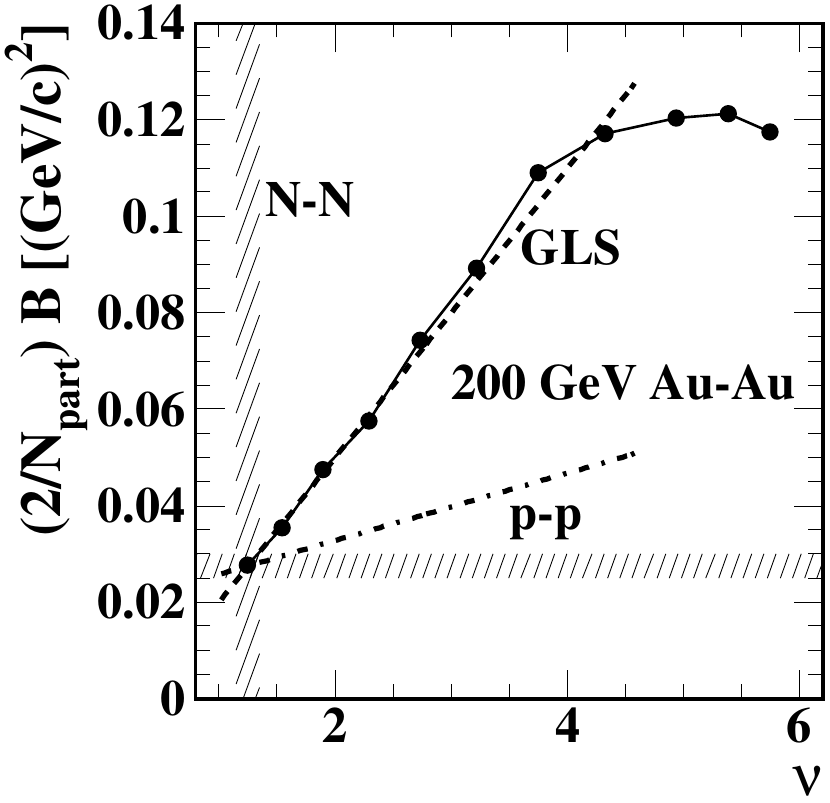}
\caption{\label{xxx}
Per-particle data from Fig.~\ref{yyy} multiplied by factor $2n_{ch}/N_{part}$ to obtain per-participant trends for \aa\ collisions equivalent to $B / n_s$ in Fig.~\ref{pp2} (right) for \pp\ collisions. The dash-dotted lines represent TCM ``first-hit'' trends if \aa\ $B_h'$ in Eq.~(\ref{pbgls}) is replaced by in-vacuum \pp\ $B_h$.
} 
\end{figure}

Figure~\ref{xxx} (left) shows 2.76 TeV \pbpb\ \pt\ fluctuation data from Fig.~\ref{yyy} (left) converted to the per-participant form vs path-length $\nu$. For $\nu < 2.5$ we observe centrality scaling $\propto \nu$ consistent with the GLS trend as expected for dijet production within transparent \aa\ collisions (following $N_{bin}$ scaling exactly). Above that point there is a 25\% increase relative to GLS until $\nu \approx 5$ above which the data show a reduction to 10\% below GLS. That panel can be compared directly with Fig.~\ref{pp2} (right)

The GLS trend for more-peripheral 2.76 TeV \pbpb\ data (dashed line) is described by
\bea \label{pbgls}
\frac{2}{N_{part}}B &=& B_{pp} +  B_h'(\nu - 1.25)~~~\text{GLS trend}
\\ \nonumber
 &=& 0.057 + 0.115(\nu - 1.25)~~\text{(GeV/c)$^2$},
\eea
with $B_{pp} = 0.057~\text{(GeV/c)$^2$}$ and $B_h' \approx 0.115$ (GeV/c)$^2$ derived from the data in Fig.~\ref{xxx} (left). From the 7 TeV \pp\ systematics of Sec.~\ref{ppmult}  $B_s = 0.039$ (GeV/c)$^2$ and $B_h = 0.024$ (GeV/c)$^2$. The sum $B_{pp} = 0.063$ (GeV/c)$^2$ derived from \pp\ data is consistent with intercept 0.057 (GeV/c)$^2$ from \pbpb\ data within the data uncertainties.  
The \pp\ ``first-hit'' hard component $B_h$ increases about 4.8-fold to $B_h'$ for subsequent \nn\ encounters (``wounded-nucleon'' effect). 
Equation (\ref{pbgls}) is reverse transformed to obtain \pbpb\ GLS (dashed) curves in previous figures.

The hard component $B_h'$ for secondary \nn\ collisions in \pbpb\ collisions is much larger than $B_h$ for in-vacuum \pp\ collisions. The dash-dotted line indicates the GLS trend predicted from \pp\ data. Secondary \nn\ collisions within \aa\ collisions appear to produce dijets at a much higher rate ($B_h' / B_h \approx 4.8$)  than isolated \pp\ collisions, consistent with results from the $\bar p_t$ study of Ref.~\cite{tommpt}.

Figure~\ref{xxx} (right)  shows 200 GeV  \auau\ \pt\ fluctuation data from Fig.~\ref{yyy} (right) converted to per-participant form vs mean participant path-length $\nu$. For $\nu < 3$ centrality scaling $\propto \nu$ is consistent with GLS (transparent \auau\ collisions). The GLS description for more-peripheral 200 GeV data is
\bea
\frac{2}{N_{part}}B  &=& 0.028 + 0.03(\nu - 1.25)~~\text{(GeV/c)$^2$},
\eea
with intercept $B_{pp} = 0.028~\text{(GeV/c)$^2$}$ compared to \pp\  $B_s = 0.021$ and $B_h = 0.007$ (GeV/c)$^2$ (from Sec.~\ref{ppedepp}) and $B_h' \approx 0.03$ (GeV/c)$^2$.
An accurate $B_{pp}$ value for 200 GeV \auau\ collisions is thus derived from 7 TeV \pp\ data.
Ratio $B_h' / B_h = 0.03 / 0.007 = 4.3$ for 200 GeV \auau\ compares with 4.8 for 2.76 TeV \pbpb. 

Reference~\cite{aliceptfluct} reports a ``strong reduction of fluctuations'' for more-central \pbpb\ collisions, referring to the data summarized in Fig.~\ref{pbpb1} (left) of the present study. The decrease is associated with possible ``onset of thermalization and collectivity.'' However, decreases in other plotting formats are much less as in Figs.~\ref{yyy} and \ref{xxx}. The present study suggests that most of the decrease for quantity $C' / \bar p_t'^2$ is due to the increasing dijet contribution to $\bar P_t^2$ in the denominator of equivalent ratio $B / \bar P_t^2$. 

While some fraction of the decreases relative to GLS above $\nu = 4.5$ in Fig.~\ref{xxx} might be attributed to ``jet quenching'' two other explanations are possible: (a) As noted, fluctuation amplitudes correspond to integration of underlying angular correlations including the SS 2D peak attributed to MB dijets~\cite{anomalous}. That peak becomes elongated on $\eta$ in more-central collisions. For the STAR and ALICE TPCs an increasing fraction of the SS peak extends beyond the detector $\eta$ acceptance with increasing centrality. (b) In each of the more-central \aa\ collisions many dijets appear within the ALICE or STAR TPC acceptance, resulting in overlap of dijet structure on the space $(\eta,\phi)$ and failure to fully resolve individual jets -- a source of inefficiency for fluctuation measurements. 

\subsection{\aa\ $\bf p_t$-fluctuations energy dependence}

The underlying energy dependence associated with the TCM for \pt\ fluctuations from \pp\ collisions is $n_s \propto \log(\sqrt s / \text{10 GeV})$ as reported in Ref.~\cite{jetspec2}. For example, the $n_s$ ratio factor is 2.18 for 7 TeV and 1.84 for 2.76 TeV relative to 200 GeV. The basic $n_s$ ($\propto$ participant low-$x$ gluons) energy trend permits prediction of NSD \pp\ values for $B_{pp}$ and $B_{pp} / n_s$ as summarized in Sec.~\ref{ppedepp}. 
Given the basic logarithmic $n_s(\sqrt{s})$ dependence the nominal TCM relation between \pp\ and \aa\ fluctuations is
\bea \label{bppeq}
B_{pp}(\sqrt{s}) \hspace{-.03in} &=& \hspace{-.03in} n_s b_s + \alpha n_s^2 b_h(\sqrt{s}) \hspace{-.0in} = \hspace{-.0in} B_s + B_h~\text{for \pp}
\\ \nonumber
B(\nu,\sqrt{s}) \hspace{-.03in} &=& \hspace{-.03in} B_{pp}(\sqrt{s}) + B_h'(\nu,\sqrt{s}) (\nu - 1.25)~\text{for \aa},
\eea
with $n_s$ in the first line given the NSD value and $\alpha \approx 0.006$.
For proper comparisons the observed \nch\ must be related accurately to inferred $n_s$ and $\nu$ as described in Secs.~\ref{hadroprod} and \ref{primes}. From Ref.~\cite{powerlaw} the predicted $\nu$ value corresponding to NSD \pp\ or \nn\ averaged over an \aa\ collision is 1.25 as in Eq.~(\ref{bppeq}).
In  that context I consider the energy dependence of \pt\ fluctuations in \aa\ collisions where there are two issues: (a) comparison of $B_{pp}$ from \pp\ collisions to an equivalent value for peripheral \aa\ collisions and (b) the energy dependence of $B'_h$ for \aa\ collisions in contrast to $B_h$ for in-vacuum \pp\ collisions.

Regarding point (a), results in Figs.~\ref{yyy} and \ref{xxx} indicate that $B_{pp}(\sqrt{s})$ values (hatched bands) predicted for two lower energies from 7 TeV \pp\ data agree with the corresponding $B(\nu,\sqrt{s_{NN}})$ (evaluated at $\nu = 1.25$) within the data uncertainties. The comparison was made without adjustment of fluctuation data or \aa\ centrality measures. The \pt\ fluctuation data  were obtained by two collaborations with different detectors and methods. The \pp\ extrapolation was made assuming $b_h(\sqrt{s})$ is independent of energy, although there is reason to expect a monotonic relation similar to that for $\bar p_{t,h}$ from Ref.~\cite{tommpt}. The $B_h$ estimates for the lower energies should then be too large, but the relative effect on the $B_{pp}$ estimate is minor because of the quadratic decrease of $B_h$ with NSD $n_s$.


Regarding point (b),  in Fig.~\ref{xxx}  $B_h' = 0.115$ and 0.03 respectively for 2.76 TeV and 200 GeV with ratio 3.8. The corresponding $\log(\sqrt{s})$ ratio is $(1.84)^2 \approx 3.4$. If $B_h' \propto B_h$ for both energies we would expect the $b_h$ ratio to be $3.8/3.4 \approx 1.1$, consistent with only slight variation of $b_h(\sqrt{s})$ with energy where I have assumed none. That result is also consistent with approximately the same ratio $B_h' / B_h \approx 4.5$ at both 2.76 TeV and 200 GeV. The ratio itself remains unexplained.

The collision-energy systematics for \pt\ fluctuations thus strongly suggest that almost all hadron production arises from participant low-$x$ gluons following a simple QCD logarithmic energy trend for $n_s$, either directly as the soft component or via large-angle scattering to dijets as the hard component. Yields, spectra, fluctuations and correlations are described quantitatively by the TCM. Whereas \pt\ fluctuation data from \pp\ collisions and more-peripheral \aa\ collisions follow a LS or GLS TCM reference accurately, the jet-related data from more-central \aa\ collisions deviate from the GLS reference quantitatively and may provide insight on \aa\ jet modifications.



\subsection{$\bf Pb$-$\bf Pb$ $\bf p_t$ fluctuations vs Monte Carlos}

In Ref.~\cite{aliceptfluct} two theory Monte Carlos are compared with the \pbpb\ data. HIJING~\cite{hijing} is an \aa\ model based on PYTHIA~\cite{pythia} (initial-state strings and minijets) plus Glauber linear superposition. AMPT~\cite{ampt} is a transport Monte Carlo based on HIJING initial conditions plus parton and hadron rescattering in the default version. In the string-melting version initial-state partons are combined via coalescence to form  hadrons.

\begin{figure}[h]
\includegraphics[width=1.6in,height=1.61in]{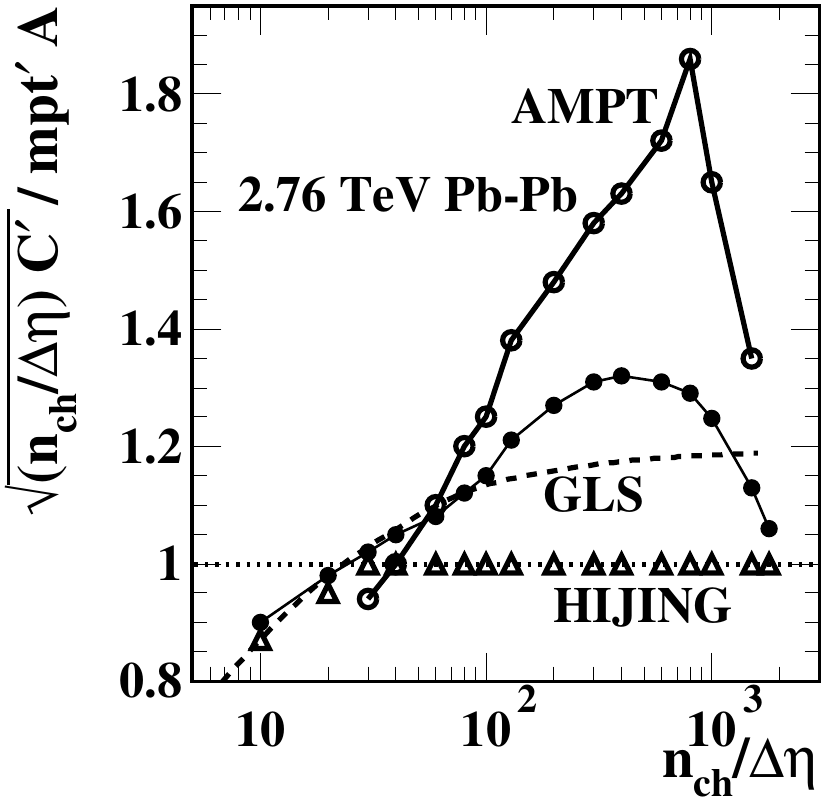}
\includegraphics[width=1.68in,height=1.65in]{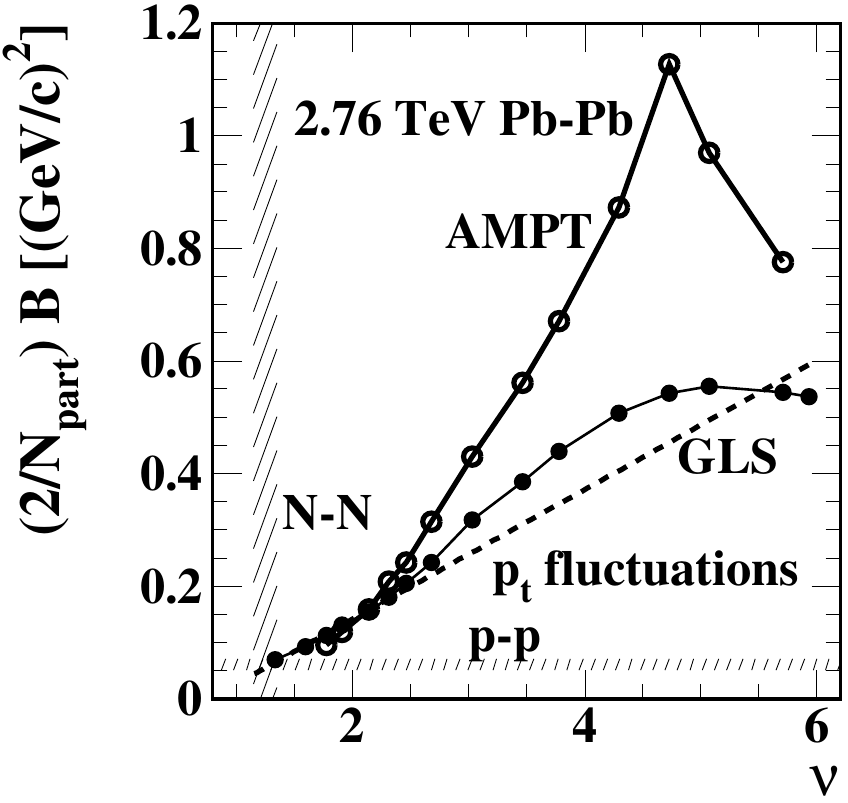}
\caption{\label{mc1}
Left: Representation of \pbpb\ data (solid points) from Fig.~8 of Ref.~\cite{aliceptfluct} including default AMPT (open circles) and HIJING (open triangles) Monte Carlo results.
Right: The \pbpb\ data and AMPT MC results transformed to per-participant format $(2/N_{part}) B$.
} 
\end{figure}

Figure~\ref{mc1} (left) repeats \pbpb\ data and results for the two Monte Carlos from Fig.~8 of Ref.~\cite{aliceptfluct}. It is noted that  HIJING data in the form $\sqrt{(n_{ch}/\Delta \eta)C'} / \bar p_t'$ is essentially constant with $A \approx 0.3$ except for the most-peripheral points. Default AMPT increases much more rapidly than the \pbpb\ data, whereas ``string melting'' AMPT increases much more slowly than the data. The HIJING results are discussed further below.

Figure~\ref{mc1} (right) shows the same data in the form $(2/N_{part}) B$ measuring the $P_t|n$ variance difference per participant pair. Default AMPT rises to twice the \pbpb\ data  for more-central collisions. Since AMPT relies on HIJING for its initial conditions, and HIJING shows negligible centrality dependence (see discussion below), one can ask what rescattering process can produce such large \pt\ fluctuations. The dashed curve represents $N_{bin}$ scaling of dijet production, the principal QCD mechanism for transporting longitudinal projectile momentum to transverse phase space in elementary collisions. That the \pbpb\ data exceed that level by 25\% in more-central collisions is already notable. For reasons given below HIJING transport is much less than required by the data. It is not clear how AMPT compensates for that deficiency.

With ``string melting'' enabled the AMPT fluctuation amplitude is much smaller than the \pbpb\ data, presumably for the same HIJING input. What happens to the transverse momentum manifested by  the default version? Since HIJING was formulated to model minijet production as described quantitatively by QCD theory why isn't dijet production a basis for discussion of results?

\begin{figure}[h]
\includegraphics[width=1.63in]{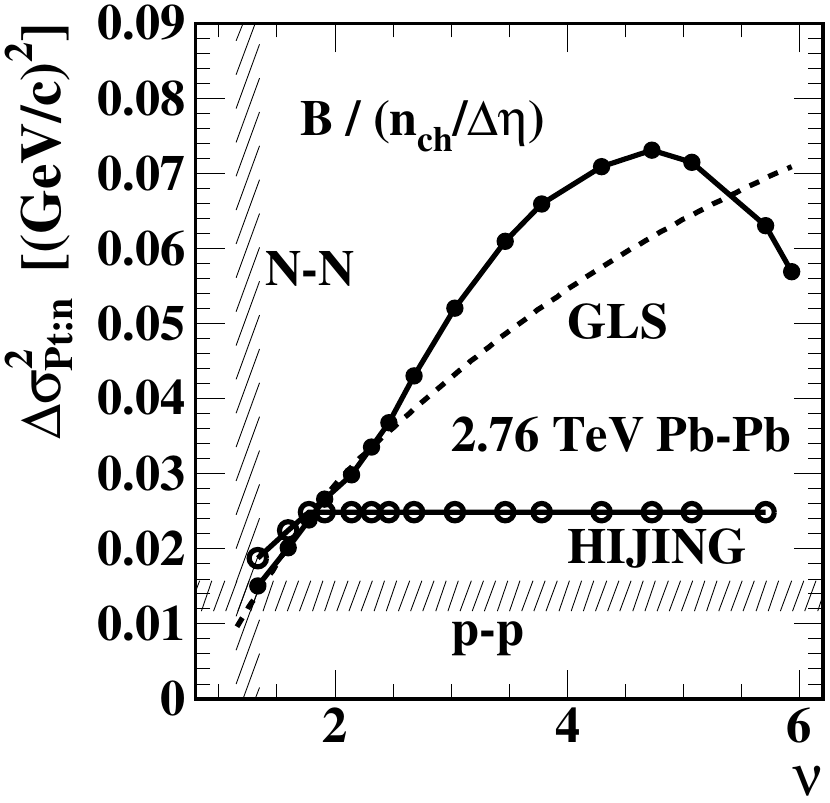}
\includegraphics[width=1.65in]{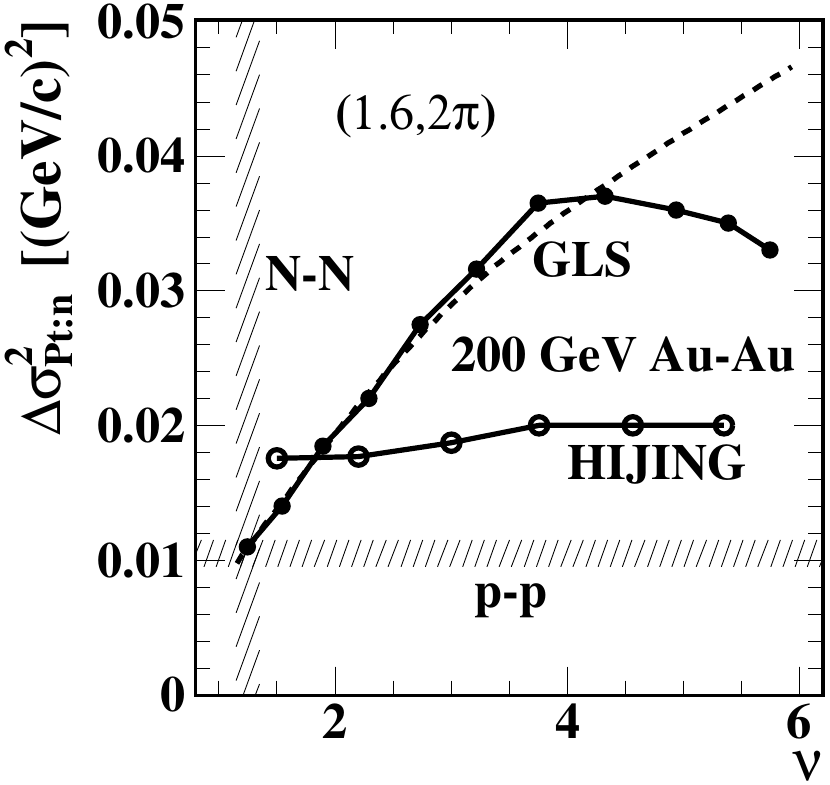}
\caption{\label{mc2}
Data and trends from Fig.~\ref{yyy} compared to HIJING MC results. The MC results in the left panel are transformed from Fig.~\ref{mc1} (left) assuming a fixed value $\bar p_t' = 0.5$ GeV/c. The MC results in the right panel are obtained from Ref.~\cite{hijscale}.
} 
\end{figure}

Figure~\ref{mc2} (left) shows 2.76 TeV \pbpb\ fluctuation data in the form $B / (n_{ch} / \Delta \eta) \equiv \Delta \sigma^2_{P_t|n}$ from Fig.~\ref{yyy} (left). The HIJING trend (open circles) is obtained from the data summary in Fig.~\ref{mc1} (left, open triangles) by first squaring those data then multiplying by $A^2 = 0.315^2$ and fixed $\bar p_t'^2 = 0.5^2$ (GeV/c)$^2$ assuming negligible $\bar p_t$ variation. 

Figure~\ref{mc2} (right) shows 200 GeV \auau\ data (solid points) from Fig.~\ref{yyy} (right) reported in Ref.~\cite{ptscale}. The open points are from a scale (bin-size) analysis of \pt\ fluctuations (see Sec.~\ref{scale}) from HIJING for 200 GeV \auau\ collisions~\cite{hijscale} corresponding to the ALICE detector acceptance ($1.6,2\pi$). The quantitative relation to the 2.76 TeV HIJING prediction is notable: 
HIJING $B / n_{ch}$ energy variation follows the data $B_{pp} / n_s$ trend.

In an earlier study it was determined that  \pt\ fluctuations from HIJING are dominated by dijet production~\cite{liu}. Disabling dijets in HIJING produced a five-fold reduction in $\Delta \sigma_{P_t|n}$ for 0-5\% central \auau\ collisions equivalent to an eighteen-fold reduction in $B$. Related angular-correlation studies show that HIJING correlation structure above 0.5 GeV/c is negligible with jets disabled. We conclude that almost all \pt\ fluctuations from HIJING arise from (mini)jets.

An explanation for the HIJING fluctuation centrality trend is provided in Sec.\ VIII I of Ref.~\cite{anomalous} relating to per-particle amplitude variation of the jet-related SS 2D peak in 200 GeV \auau\ number (as opposed to \pt) angular correlations. The HIJING centrality trend is nearly constant while the \auau\ data trend shows an eight-fold increase from the \pp\ value. The difference arises from two sources. The HIJING MC produces a hard-component (jet-related, binary-collision scaling) hadron yield per binary collision that is seven times the \pp\ yield and 1.6 times the more-central \auau\ yield, but the number of jet-correlated hadron pairs is 60\% of the \pp\ equivalent and only 20\% of the more-central \auau\ equivalent. Thus, a {\em per-particle} correlation measure such as $\Delta \sigma^2_{P_t|n}$ (number of correlated pairs / number of particles) may increase rapidly for \auau\ data but not at all for HIJING data. The same trends apparently determine \pt\ fluctuations for 200 GeV and 2.76 TeV as in Fig.~\ref{mc2}. 

The underlying difference is the PYTHIA model~\cite{pythia} for dijet production in \pp\ collisions that assumes an eikonal approximation inconsistent with \pp\ spectrum data~\cite{tommpt} and a default jet spectrum lower limit $p_0 = 2$ GeV for HIJING compared to the observed 3 GeV~\cite{jetspec2}, resulting in a large excess of very-low-energy partons.






\section{$\bf p_t$ Fluctuations $\bf vs$ correlations} \label{scale}

The LHC \pt\ fluctuation measurements correspond to a single 2D $(\eta,\phi)$ bin size (scale) -- the TPC detector acceptance. In previous studies the scale variation of \pt\ fluctuations was measured for \auau\ collisions and directly related to underlying angular correlations~\cite{inverse,hijscale,ptscale}.  To establish a physical interpretation for event-wise \pt\ fluctuations at the LHC it is useful to review results from published fluctuation scaling studies at the RHIC.

\subsection{Total-variance scaling and fluctuation inversion}

An early \pt\ fluctuation study of RHIC data reported in Ref.~\cite{jeffptfluct} related the concept of {\em total variance} introduced in Ref.~\cite{clt} to analysis of fluctuation scale dependence. The ensemble-mean total variance for event-wise \pt\ and particle-number distributions on space $x$ with $M$ bins of scale (bin width) $\delta x$ within acceptance $\Delta x$ is
\bea
\Sigma^2_{P_t|n}(\Delta x,\delta x) &=& \sum_{a=1}^M \overline{[P_{t}(\delta x) - n(\delta x) \bar p_t]^2_{a}},
 \eea
where $P_{t}(\delta x)$ and $n(\delta x)$ are bin sums and $n_{ch}$ is the multiplicity in acceptance $\Delta x$.
The CLT is equivalent to the statement that for certain conditions (independent samples from a fixed parent process) the total variance is scale invariant. In general the total variance approaches the limit $n_{ch} \sigma^2_{p_t}$ at the ``single-particle'' scale [$\delta x \ll \Delta x$, one particle per bin in $M(\Delta x) = n_{ch}$ occupied bins]. The total-variance difference $\Delta \Sigma^2_{P_t|n}(\delta x_1,\delta x_2)$ over some scale interval $[\delta x_1,\delta x_2]$ is nonzero if CLT conditions are not met: The parent process varies from event to event and/or the samples (e.g.\ particle momenta) are correlated. Quantity $B$ defined in Eq.~(\ref{bbar}) is the $P_t|n$ total-variance difference evaluated over the maximum accessible scale interval -- between the detector-acceptance scale and the single-particle scale.

\begin{figure}[h]
\begin{tabular}{cc}
\begin{minipage}{.47\linewidth}
\end{minipage} &
\begin{minipage}{.47\linewidth}
\end{minipage}\\
\begin{minipage}{.47\linewidth}
\includegraphics[keepaspectratio,width=1.65in]{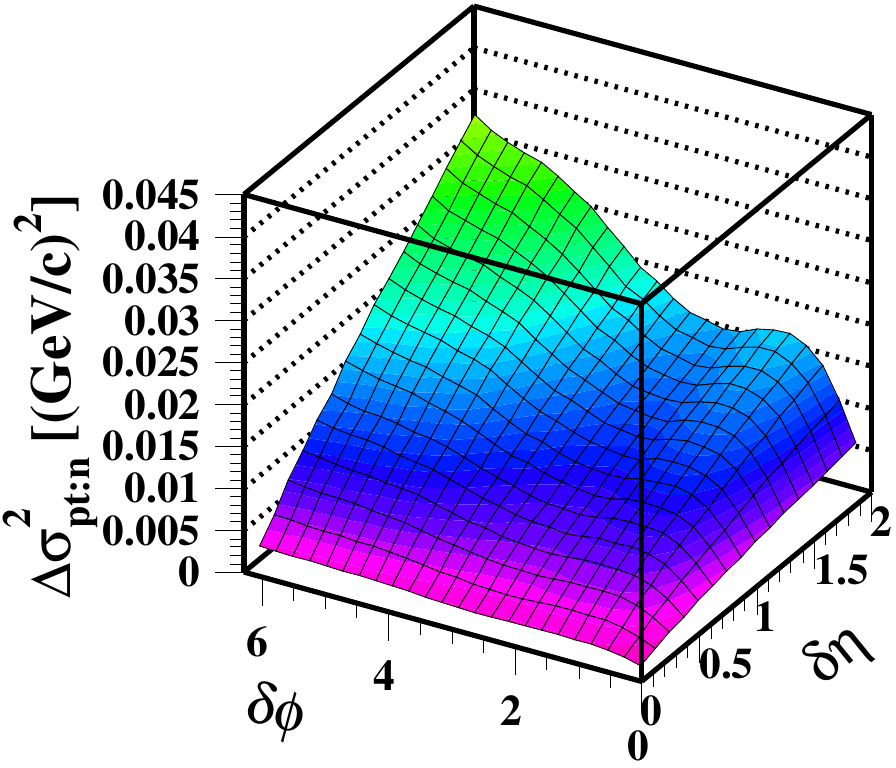}
\end{minipage} &
\begin{minipage}{.47\linewidth}
\includegraphics[keepaspectratio,width=1.65in]{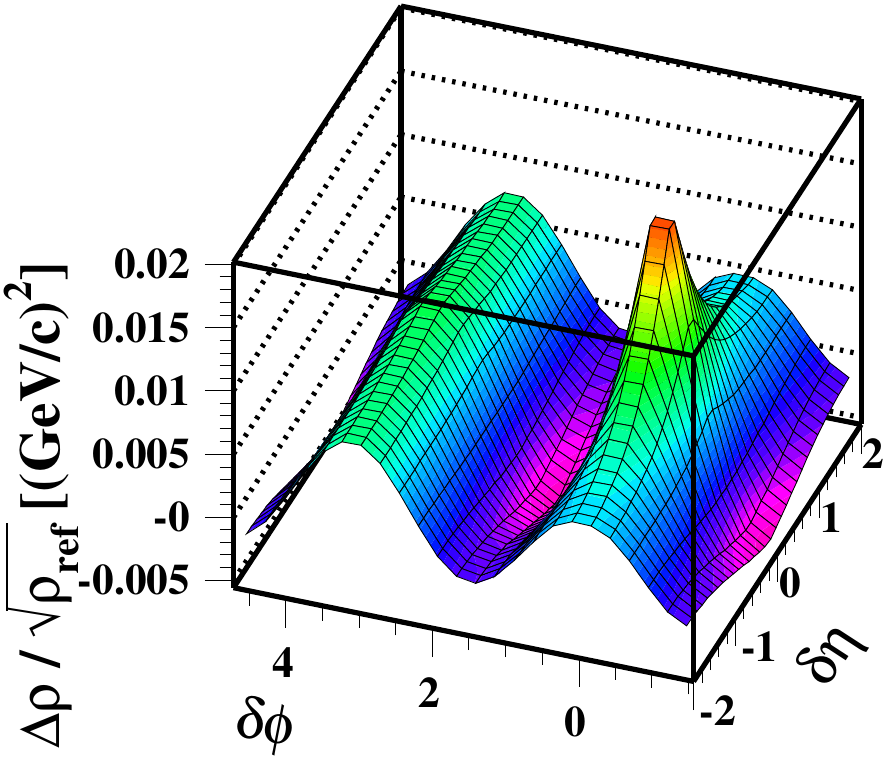}
\end{minipage}\\
\begin{minipage}{.47\linewidth}
\end{minipage} &
\begin{minipage}{.47\linewidth}
\end{minipage}\\
\end{tabular}
\caption{
(Color online) \label{fluctinv}
Left: $\Delta \sigma^2_{p_{t}:n}$ (GeV/c)$^2$ distributions on scale $(\delta \eta,\delta \phi$) for 45-55\% central 200 GeV \auau\ collisions. 
Right: Corresponding autocorrelations on difference variables ($\eta_{\Delta},\phi_{\Delta}$) inferred from data at left by fluctuation inversion.
}  
\end{figure}

Figure~\ref{fluctinv} (left) shows the 2D scale dependence of {\em per-particle} variance difference (relative to the single-particle scale) $\Delta \sigma^2_{P_t|n}(\delta \eta,\delta \phi) = \Delta \Sigma^2_{P_t|n}(\delta \eta,\delta \phi) /\bar n$ within the STAR TPC acceptance for 200 GeV \auau\ collisions~\cite{ptscale}. The ALICE \pbpb\ fluctuation data reported in Ref.~\cite{aliceptfluct} correspond to a single point at $(1.6,2\pi)$ on a similar surface representing 2.76 TeV \pbpb\ collisions. The scale variation of $\Delta \Sigma^2_{P_r|n}(\delta \eta,\delta \phi)$ has been expressed as the running integral of a 2D angular autocorrelation on difference variables $(\eta_\Delta,\phi_\Delta)$ in the form of an integral equation. The underlying angular correlations can be inferred by inverting that integral equation~\cite{inverse}.

Figure~\ref{fluctinv} (right) shows the inferred $p_t$ angular correlations~\cite{ptscale}. The result has three basic elements: (a) a same-side (SS) 2D peak, (b) an away-side (AS) 1D peak on azimuth and (c) a non-jet cylindrical quadrupole [$\cos(2\phi_\Delta)$ dependence]. The general combination is the same as that observed for {\em number} angular correlations~\cite{axialci,anomalous}, but there are quantitative differences in the SS 2D peak structure. Elements (a) and (b) have been identified with dijet production in a number of ways~\cite{porter2,porter3,anomalous,hardspec,fragevo,jetspec}. Element (c) might be related to elliptic flow if that were relevant to nuclear collisions~\cite{davidhq,nohydro,quadspec,noelliptic}.

\subsection{$\bf p_t$ correlations vs $\bf p_t$ fluctuations} \label{compare}

\begin{figure}[h]
\includegraphics[keepaspectratio,width=1.65in]{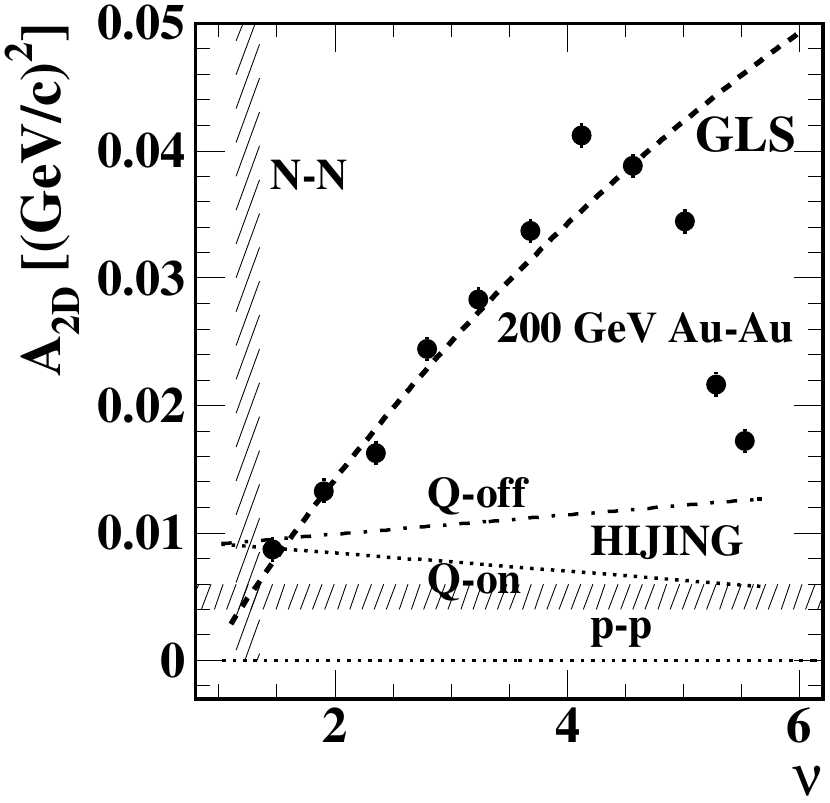}
\includegraphics[keepaspectratio,width=1.65in]{alice11g}
\caption{ \label{fits}
Left: Jet-related same-side 2D peak amplitude (solid dots) vs path length $\nu$ from 2D fits to \pt\ angular correlations from 200 GeV \auau\ collisions as in Fig.~\ref{fluctinv} (right).  The dash-dotted and dotted lines represent corresponding HIJING quench-off and quench-on results respectively from Ref.~\cite{hijscale}. There is no soft component to those jet-related data.
Right: Per-particle \pt\ fluctuation amplitude $B / n_{ch}$ as shown in Fig.~\ref{yyy} (right) for comparison with correlations, including the \pp\ estimate $B_{pp} / n_{s} \approx  0.0112$ (GeV/c)$^2$ from Sec.~\ref{ppedepp}.
} 
\end{figure}

Figure~\ref{fits} (left) shows the best-fit SS 2D peak amplitude $A_{2D}$ (solid points) for 200 GeV \auau\ data {\em vs} path length $\nu$~\cite{ptscale}. The peak amplitude increases with centrality to a maximum value and then decreases for the most central collisions. The monotonic increase for more-peripheral collisions, approximately proportional to $\nu$ (dashed curve), is consistent with the binary-collision scaling expected for dijet production. $A_{2D}$ is closely correlated with the amplitude of the AS-dipole component of the 2D fit model, as expected for dijet correlations.  

Figure~\ref{fits} (right) shows  \pt\ fluctuations from 200 GeV \auau\ collisions as  the integral (up to a specific bin size or scale) of \pt\ angular correlations evaluated at the ALICE TPC angular acceptance~\cite{ptscale}. The two panels are directly related since angular correlations in the left panel are obtained by inversion of the scale dependence of fluctuations as represented in the right panel. 

Note that for 200 GeV \pt\ fluctuations and correlations there is no sign of the  ``sharp transition'' appearing near $\nu = 3$ in number angular correlations reported in Ref.~\cite{anomalous}. The substantial increase in jet-correlated hadron pairs above the transition point was attributed to strong modification of parton fragmentation in the \aa\ environment~\cite{fragevo}. The same study indicated that modified jets still retain almost all of the parton energy and hence most jet-related \pt, consistent with Fig.~\ref{fits}.



\subsection{\auau\ $\bf p_t$ angular correlations} \label{ptangle}

\pt\ angular correlations can be obtained either by inversion of the scale dependence of \pt\ fluctuations~\cite{ptscale,inverse} or by direct pair counting. The same jet-related correlation structures are observed, with minor quantitative differences in the SS 2D peak structure. Those results indicate that the \mpt\ fluctuations expected to reveal critical fluctuations of temperature near a QCD phase boundary are actually dominated by a MB jet (minijet) contribution~\cite{axialci,anomalous,hardspec,fragevo,ptscale}. 

\begin{figure}[h]
  \includegraphics[width=1.65in]{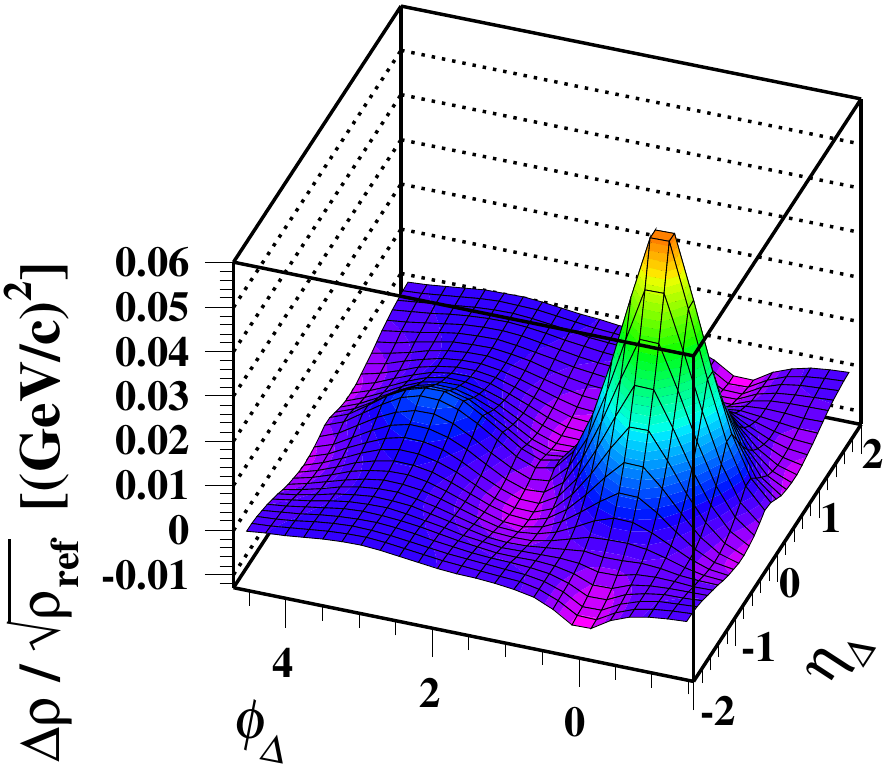}
  \put(-95,90) {\bf (a)}
  \includegraphics[width=1.65in]{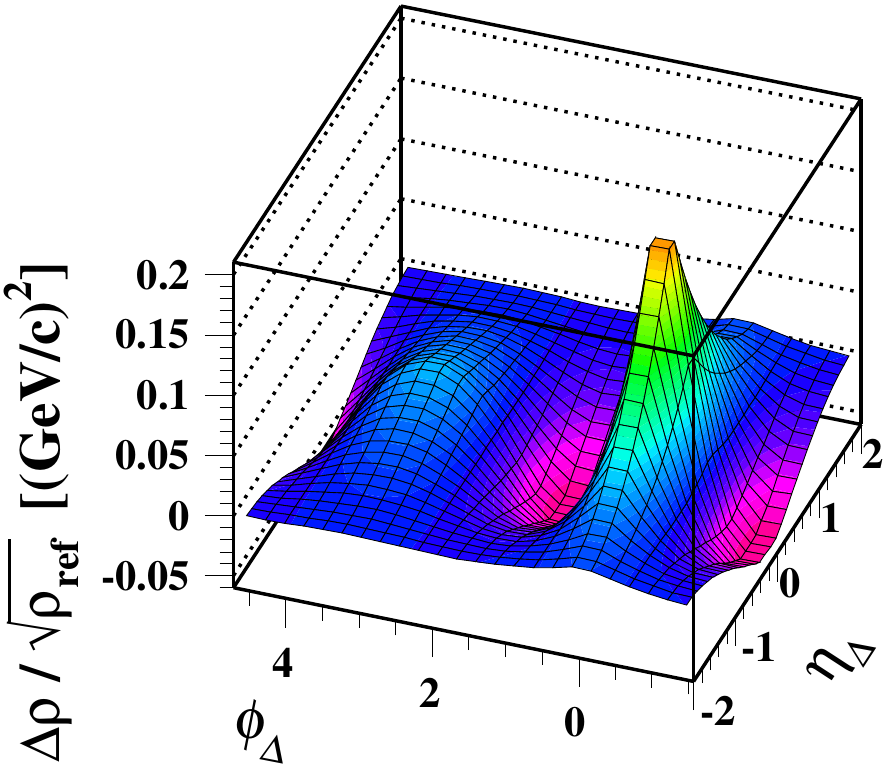}
  \put(-95,90) {\bf (b)}  \\
  \includegraphics[width=1.65in]{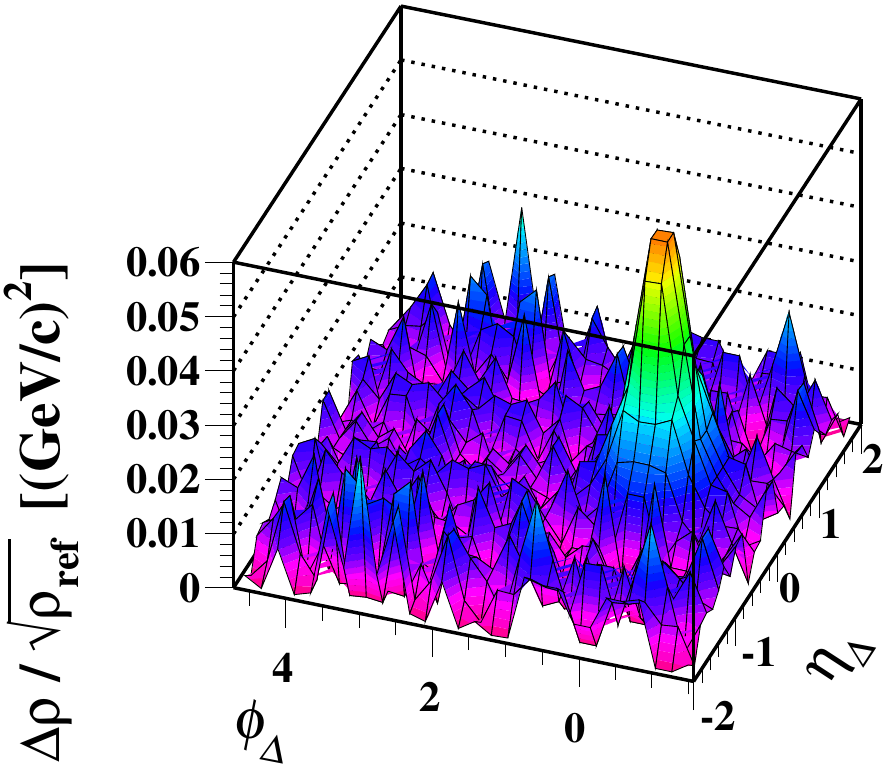}
  \put(-95,90) {\bf (c)}  
  \includegraphics[width=1.65in]{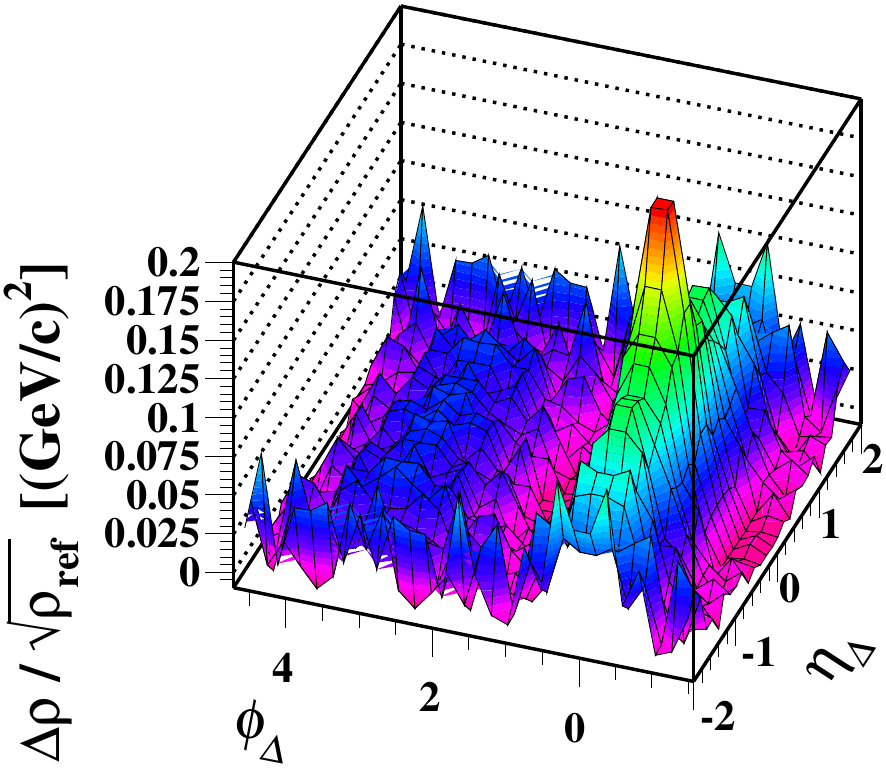}
  \put(-95,90) {\bf (d)}\caption{\label{ptscale} (Color online)
Upper: \pt\ angular correlations for  (a) 85-95\% and (b) 10-20\% central 200 GeV \auau\ collisions inferred by inverting \pt\ fluctuation scale dependence~\cite{ptscale}. AS dipole and nonjet quadrupole components of 2D model fits to the data are subtracted.
Lower:  Results for the same collision systems but \pt\ correlations are obtained by direct pair counting rather than fluctuation inversion. Improved angular resolution and unfiltered statistical fluctuations are evident. Data for all panels include an additional acceptance factor $4\pi$.
 }  
 \end{figure}

Figure~\ref{ptscale} (upper panels) shows \pt\ angular correlations for  (a) 85-95\% and (b) 10-20\% central 200 GeV \auau\ collisions obtained by inversion of \pt\ fluctuation scale dependence~\cite{inverse}. Fitted AS dipole and nonjet quadrupole components have been subtracted to isolate the SS 2D peak structure. Similar analysis of HIJING Monte Carlo data supports a jet interpretation for the SS peak~\cite{hijscale}.

Figure~\ref{ptscale} (lower panels) shows  \pt\ angular correlations for the same collision systems obtained by direct pair counting, confirming the results in the upper panels obtained by fluctuation scale inversion. The SS 2D peak for \pt\ correlations is narrower than that observed for number correlations. The difference is expected for jet correlations, since fewer fragments with larger momenta are found closer to the jet thrust axis and more fragments  with smaller momenta appear at larger angles.

\vskip .1in

From these comparisons between \pbpb\ and \auau\ \pt\ fluctuations and between \auau\ \pt\ fluctuations and jet-related \pt\ angular correlations we may conclude that there are strong similarities between the 2.76 TeV \pbpb\ \pt\ fluctuation data and 200 GeV \auau\ \pt\ {\em and} number angular-correlation data identified with dijet production. Per-participant measures typically exhibit nearly linear increase $\propto \nu$ (GLS) for more-peripheral collisions possibly followed by an interval of significant increase above GLS (much larger for number than for \pt\ correlations) and significant decrease for most-central collisions. The GLS trend for \pt\ fluctuations is consistent with dijet production (binary-collision scaling) and with jet-related structure in both number and \pt\ angular correlations. 



\section{Discussion} \label{disc}

\subsection{LHC fluctuation analysis and interpretations}

The analysis in Ref.~\cite{aliceptfluct} is similar to several previous studies at the SPS and RHIC motivated by a search for critical fluctuations associated with the structure of the QCD phase boundary. It is conjectured  that event-wise mean \pt\ may represent the temperature of a thermodynamic state formed in \aa\ collisions, in which case critical temperature fluctuations near a QCD phase boundary or critical point may be reflected in mean-\pt\ fluctuations.

Given the definition of quantity $C$ in  Eq.~(\ref{aliceptfluctx}) as a measure of ``nonstatistical'' $\langle p_t \rangle$ fluctuations (excess variance relative to a reference) and the context of the temperature narrative $\sqrt{C} \sim \delta T$ represents an r.m.s.\ measure of excess temperature fluctuations and $\bar p_t \sim T_0$. The adopted fluctuation measure $\sqrt{C} / \bar p_t$ then emulates $\delta T / T_0$ as an r.m.s.\ measure of relative temperature fluctuations.

It is concluded that there is no significant energy dependence of \mpt\ fluctuations for \pp\ or \aa\ collisions over a large energy interval. The \pp\ data appear to decrease according to a power-law trend on \nch. The \aa\ data decrease with centrality (and therefore \nch) but with substantial deviations from the \pp\ power-law trend. The decrease with \aa\ centrality is said to be expected for a ``dilution scenario caused by [linear] superposition of partially independent particle-emitting sources.'' The linear-superposition reference is assumed to be $1/\sqrt{n_{ch}}$.
Deviations of \pbpb\ fluctuations from  the LS reference are said to be ``consistent with'' (a) string percolation or (b) onset of thermalization and collectivity. No ``critical behavior'' is observed (but the concept is not defined). Initial-state density fluctuations may also contribute.

The analysis and interpretation of Ref.~\cite{aliceptfluct} rely on a number of assumptions that may be questioned. It is assumed that a thermodynamic state with a well-defined temperature may be formed in high-energy nucleus-nucleus collisions and \mpt\ is a temperature estimator. But when isn't a thermodynamic state formed: peripheral collisions, \pp\ collisions?  If a phase transition were relevant to high-energy nuclear collisions  then some related fluctuation excess might arise. But the assumption that ``nonstatistical fluctuations'' in some statistical measure may necessarily reveal a phase transition is not justified.
Linear superposition of independent subsystems is said to result in reduction (dilution) of relative  temperature fluctuations. But if most subsystems down to individual \nn\ collisions are independent how does the composite system qualify as a thermodynamic state? 

Use of \pp\ \mpt\ fluctuations as a baseline or reference for \aa\ collisions is proposed but not implemented. Possible ``conventional mechanisms'' for \mpt\ fluctuations in \pp\ collisions are acknowledged, including jets, but are not pursued further.  
It is often assumed that jets do not contribute to $p_t < 2$ GeV/c in \aa\ collisions~\cite{nojets} (possibly what motivates the $p_t < 2$ GeV/c cut for the LHC analysis), but there is plentiful evidence that {\em most} jet-related hadrons appear below 2 GeV/c and within resolved jets~\cite{eeprd,jetspec}.
Analysis of LHC $\bar p_t$ systematics~\cite{tommpt} shows that variation of $\bar p_t$ with \nch\ is dominated by dijet production, contrary to some assumptions of Ref.~\cite{aliceptfluct}.

Previous \pt\ fluctuation analyses at the RHIC are cited (in Ref.~\cite{aliceptfluct} as [15-20]) but only [18] appears again in the text -- motivating measure $C$ in Eq.~(2).  Angular-correlation and energy-dependence results from [19,20] revealing a dijet contribution dominating \pt\ fluctuations and angular correlations are not mentioned. Although Refs. [18] and [20] disagree diametrically about \pt\ fluctuation energy dependence that is not acknowledged. 

Certain Monte Carlo models are said to be in ``qualitative agreement'' with the fluctuation data, but that implies quantitative {\em dis}\,agreement to an extent that may falsify the models.  HIJING is assumed to represent linear superposition of N-N collisions (but see Sec.~\ref{linsup}), and any difference from HIJING data should then indicate mechanisms unique to A-A collisions. But HIJING based on PYTHIA includes an incorrect model for \pp\ collisions, as noted in Refs.~\cite{anomalous,jetspec2}, and deviates dramatically from an \aa\ GLS reference {\em based on measurements} that represents superposition of real  \nn\ collisions.



\subsection{Fluctuation measure definitions}

Several \pt\ fluctuation measures applied to data at the SPS, RHIC and LHC are reviewed in Sec.~\ref{fluctsum}. The measure definitions appear to follow two opposing principles: (a) a temperature narrative motivating model-dependent ``ratios of ratios'' (intensive quantities) in which cancellations may conceal significant systematic trends and (b) a model-independent approach based on extensive quantities (e.g.\ $n_{ch}$, $P_t$) and their statistical properties. The primary event-wise RVs are total $P_t$ and total $n_{ch}$ within a single fixed angular acceptance or within each of several angular bins covering that acceptance.  Deliberate choices must be made regarding combinations of those RVs to form secondary fluctuation measures.

Reference~\cite{aliceptfluct} applies the term ``fluctuation strength'' to $\sqrt{C} / \bar p_t$, but the term could be applied as well to several other statistical measures that incorporate $n_{ch}$ and $P_t$ as RVs including variance difference $B$. Both $C$ and $\bar p_t$ are ratios of mean values then combined to form a secondary ratio. Large but similar fluctuations in the primary RVs may nearly cancel in the square root of a ratio of ratios, possibly obscuring significant collision mechanisms. 

One should first measure fluctuation trends for each primary RV separately (variances) and in combinations (covariances) relative to well-defined statistical references (representing CLT conditions). The choice of fluctuation measure should be compatible with correlation measures given the direct algebraic relation~\cite{inverse}. Data presentation involves both y-axis (fluctuation measures) and x-axis (system ``size'' measures $n_{ch}$, $N_{part}$, $\nu$) choices. A good plotting format may test a significant hypothesis (e.g.\ Fig.~\ref{pp2} right, Fig.~\ref{xxx}) while a poor format may obscure an important data trend (e.g.\ Fig.~\ref{pp1} left, Fig.~\ref{pbpb1} right).

\subsection{Linear-superposition references} \label{linsup}


Reference~\cite{aliceptfluct} defines an ``independent superposition'' (of unspecified particle sources)  reference as $\sqrt{C} / \bar p_t \propto 1/\sqrt{n_{ch}}$ or $n_{ch} C /  \bar p_t^2 \propto$ constant. But $n_{ch} C \approx \Delta \sigma^2_{P_t|n}$ and the ratio $\Delta \sigma^2_{P_t|n} /  \bar p_t^2$ would remain independent of \aa\ centrality only if the numerator and denominator happen to vary in proportion or each remains constant. The latter {\em would} hold in the absence of dijet production. The quantities would then include only soft components with $\Delta \sigma^2_{P_t|n} \rightarrow B_s / n_s$ and $ \bar p_t \rightarrow \bar p_{t,s}$. The defined reference is thus equivalent to claiming that dijet production does not contribute significantly to \mpt\ fluctuations.

Linear (independent) superposition of particle sources is the basic assumption of LS references for the TCM in which (at least) two specific hadron sources are assumed to contribute: The soft component represents projectile-nucleon dissociation (longitudinal fragmentation) and the hard component represents  transverse fragmentation of large-angle-scattered parton pairs to dijets. 

For a LS reference jet-related contributions to yields and spectra~\cite{ppprd,hardspec} as well as to fluctuations and correlations should scale with \pp\ multiplicity as $n_s^2$ assuming linear superposition of gluon-gluon binary encounters (e.g.\ Fig.~\ref{pp2}, left) and with \aa\  centrality as $N_{bin}$ assuming a GLS reference for transparent \aa\ collisions. Per-participant measures should then scale linearly with $n_s$ for \pp\ collisions (e.g.\ Fig.~\ref{pp2}, right) and with $\nu$ for \aa\ collisions (e.g.\ Fig.~\ref{xxx}). A MB dijet (average over the MB dijet spectrum at given energy) is observed to contribute a certain fixed amount to $\bar p_t$~\cite{tommpt}, to \pt\ fluctuations $B$ and to number and \pt\ angular correlations on $(\eta,\phi)$~\cite{ptscale}. 

Because each of \nch, $P_t$, $B$ and jet-related spectrum and correlation components has a unique TCM representation (but with similar forms) an LS reference for any one quantity cannot be simply expressed in terms of another, as proposed in Ref.~\cite{aliceptfluct}. It is only in  the limit of no dijet contribution that the surviving soft components would be simply related, with $n_{ch}C \propto $ constant and $\bar p_t$ independent of \pp\ multiplicity or \aa\ centrality. 
HIJING with jet production enabled and no jet quenching follows that trend approximately, but detailed study of HIJING yields and correlations~\cite{anomalous} reveals that HIJING overpredicts the hard-component multiplicity while underpredicting hard-component jet correlations (relative to \aa\ data) such that the ratio $B / n_{ch} \approx n_{ch}C$ is nearly independent of \aa\ centrality as in Fig.~\ref{mc2}. HIJING is thus not representative of GLS scaling in \aa\ collisions.

\subsection{The dominant role of minimum-bias dijets}

Reference~\cite{aliceptfluct} does acknowledge jets as a possible ``conventional'' mechanism for particle production. It is implied that jet contributions to \pp\ collisions might then be used to identify equivalent structure in \aa\ collisions, but the only comparison of \pp\ and \pbpb\ structure in that study is Fig.~5 (or equivalently Fig.~8) where no jet contribution is identified.

 In the TCM context a strong dijet contribution to LHC \pt\ fluctuation data is easily identified  for both \pp\ and \pbpb\ collisions. Figure~\ref{pp2} shows variance difference $B$ varying with $n_s$ in exact accord with the \pp\ TCM LS reference within data uncertainties. The hard component (dijets) is observed to dominate \pp\ \pt\ fluctuations, varying in proportion to $n_s^2$ representing gluon-gluon binary collisions. In Fig.~\ref{xxx} (left) data for more-peripheral \pbpb\ collisions again show agreement with GLS scaling expected for dijet production in \aa\ collisions. With increasing \aa\ centrality measure $B$ exceeds $N_{bin}$ scaling by about 25\%, consistent with persistence of the dijet mechanism but with some quantitative modification.

An analysis of LHC $\bar p_t$ vs \nch\ data for several energies and collision systems indicates that $\bar p_t$ data trends are all accurately described by the TCM. The \pp\ data follow LS scaling across a ten-fold increase of $n_s$. The dijet production rate then increases 100-fold, implying multiple MB dijets per \pp\ collision on average~\cite{tommpt} and consistent with LHC \pp\ number angular correlations~\cite{cmsridge,cmsridge2}.

200 GeV \auau\ \pt\ fluctuation data show trends very similar to the LHC data represented in  this study~\cite{ptscale,ptedep}. Those results are in turn consistent with number correlation measurements indicating that dijets are a dominant particle production mechanism~\cite{anomalous} and suggesting that jet manifestations are very similar at the RHIC and LHC modulo a basic QCD $\log(\sqrt{s} / \text{10 GeV})$ scale factor for $n_s$.

Previous to  the present study a wealth of evidence for dijet dominance of high-energy nuclear collisions has been presented. The MB dijet-based TCM context is internally consistent and has been employed to predict and explain many experimental results from the RHIC and LHC, including (a) systematics of LHC ensemble-mean $\bar p_t$~\cite{tommpt}, (b) \pt\ angular correlations~\cite{ptscale,ptedep}, (c) number angular correlations~\cite{axialci,porter2,porter3,anomalous}, (d) trigger-associated transverse-rapidity correlations~\cite{trigmodel,pptheory,pptrig}, (e) \pt\ spectra~\cite{ppprd,hardspec} and (f) jet-related systematics of hadron yields~\cite{ppprd,hardspec,jetspec}. Those results are all in accord with measured dijet properties~\cite{eeprd,jetspec2} and QCD theory~\cite{fragevo}.

\section{Summary}\label{summ}

A measurement of fluctuations in event-wise mean transverse momentum denoted by \mpt\ from \pp\ and \pbpb\ collisions at the large hadron collider (LHC) has been reported recently. The fluctuation measure denoted by $\sqrt{C} / \bar p_t$ is motivated by a temperature narrative in which collisions attain some degree of thermalization and are characterized by a temperature estimated by \mpt\ as one property of a thermodynamic state. Excess \mpt\ fluctuations compared to a reference might indicate the presence of a phase boundary between a conjectured quark-gluon plasma (QGP) phase and a hadron-fluid phase.

It is inferred from the LHC data that \mpt\ fluctuation ``strength'' is nearly independent of collision energy over a broad interval for both collision systems. For \pp\ collisions fluctuations are said to decrease with increasing particle multiplicity \nch\  approximately as a power law $n_{ch}^{-0.4}$. For \pbpb\ collisions \mpt\ fluctuations also decrease overall, but relative to the \pp\ trend they increase for mid-central collisions and then decrease for most-central collisions. The \pbpb\ results are said to be consistent with models that incorporate collective phenomena.

The choice of fluctuation measure  $\sqrt{C} / \bar p_t$ from among several candidates is a critical step in such data analysis. In the present study I review several measures applied previously to fluctuation analysis of nuclear collision data and describe their algebraic relationships. I identify total multiplicity \nch\ and total transverse momentum $P_t$ (falling within some detector angular acceptance) as the basic {\em extensive} random variables for the data system, with \mpt\ = $P_t / n_{ch}$ as a derived {\em intensive} ratio. I introduce {\em variance difference} $B$ for $P_t$ {\em conditional on} \nch\ as a physical-model-independent \pt\ fluctuation measure.

Given the relation $C \approx B / n_{ch}^2$, data from the LHC \mpt\ analysis can be converted to other formats and compared directly with previous analysis at the relativistic heavy ion collider (RHIC). RHIC fluctuation analyses employing a {\em per-particle} variance-difference measure in the form $B / n_{ch}$ revealed \pt\ fluctuations increasing strongly with \auau\ collision centrality and with collision energy, very different from the reported LHC trends. The scale (angle-bin-size) dependence of $B / n_{ch}$ was also measured and inverted via a standard numerical method to reveal the underlying \pt\ angular correlations. Principal features of the inferred \pt\ correlation structure were identified with minimum-bias jets and were subsequently confirmed by correlation analysis based on direct pair counting.

The LHC \pp\ \mpt\ fluctuation data, when converted to measure $B$, are described accurately by a two-component (soft+hard) model (TCM) in which the hard component represents minimum-bias (MB) dijets. The \pp\ TCM has been successful in describing yield, spectrum and correlation data at the RHIC and, most recently, ensemble-mean $\bar p_t$ vs \nch\ trends from the LHC. The TCM description of $B$ vs \nch\ for LHC \pp\ collisions confirms that MB dijets dominate \pt\ fluctuations for larger event multiplicities.

The TCM can also be applied to \aa\ collision data, with {\em Glauber linear superposition} (GLS) of nucleon-nucleon (\nn) collisions within \aa\ collisions as a reference. The TCM description of \pbpb\ \mpt\ fluctuation data converted to $B$ indicates that \pt\ fluctuations follow a GLS trend with collision centrality for more-peripheral \pbpb\ collisions (indicating transparency) but deviate quantitatively from that trend for more-central collisions.

The energy dependence of \pt\ fluctuations from \pp\ collisions measured by $B$ are predicted over a range from RHIC to LHC energies by a simple $\log(s/s_0)$ trend consistent with QCD field theory and with measured systematics of MB jet spectra. $B$ values from peripheral \pbpb\ ($\approx$ \nn) collisions are quantitatively consistent with the values from non-single-diffractive \pp\ collisions, including the energy dependence from RHIC to LHC. In that comparison $B_{pp}$ values for RHIC 200 GeV \pp\ collisions were successfully inferred from LHC 7 TeV \pp\   $\sqrt{C} / \bar p_t$ data.

The overarching message from the LHC \pt\ fluctuation data appears to be that MB dijets play a dominant role in all high-energy nuclear collisions, consistent with previous analysis of yields, spectra and correlations at the RHIC. Jet manifestations are clearly evident in extensive measures \nch\ and $P_t$ and are simply and accurately represented by the TCM over a range of collision systems and energies. Measures that rely on ratios of means such as $\bar p_t$ or ratios of random variables such as \mpt\ present an ambiguous picture because of possible cancellation of dijet contributions. Data in the form  $\sqrt{C} / \bar p_t$, a ratio of ratios including an additional factor $1/n_{ch}^2$ compared to conventional variance fluctuation measures, are difficult to interpret in that form. Transformation of such data to a variance-difference format presents a clearer picture.

This material is based upon work supported by the U.S.\ Department of Energy Office of Science, Office of Nuclear Physics under Award Number DE-FG02-97ER41020.

\begin{appendix}

\section{$\bf \bar p_t $ TCM for \aa\ collisions} \label{aatcm}

The TCM for \aa\ collisions is based on the Glauber model in which the fractional cross section (centrality) $\sigma/\sigma_0$ is related to geometry parameters $N_{part}$ the number of projectile nucleon participants, $N_{bin}$ the number of binary \nn\ encounters and $\nu = 2N_{part} / N_{bin}$ the mean participant pathlength in number of \nn\ encounters. The correspondence with observable \nch\ can be established from the MB cross-section distribution on \nch. For the present study the correspondence between \nch\ reported in Ref.~\cite{aliceptfluct} and Glauber model parameters was determined as described in Sec.~\ref{hadroprod} and Ref~\cite{tommpt}.

For \aa\ collisions the $\bar p_t$ TCM of Eq.~(\ref{pptcm}) or (\ref{pptcmp}) must be modified in three ways: (a) the multiplicity hard component increases with centrality as $n_h(\nu)$, (b) due to modified parton fragmentation in more-central \aa\ collisions the spectrum hard-component shape changes (softens) with centrality leading to variation of $\bar p_{t,h}$ as $\bar p_{t,h}(\nu)$ and (c) an \nn\ ``first encounter' effect must be accommodated, with details presented in Sec.~\ref{tcmintro}.

The direct extension of \pp\ $n_{ch} = n_s + n_h$ to  \aa\ is the first line of Eq.~(\ref{aanchx}) where the \nn\ soft and hard components are scaled up by the corresponding Glauber parameters. The observed trend for hadron production in \aa\ collisions implies that $n_h$ for the first \nn\ encounter is the same as that for \pp\ independent of the \aa\ centrality, but for $\nu - 1$ subsequent encounters $n_h$ transitions to $n_h'(\nu)$ depending on \aa\ centrality. The consequence is the second line that accurately describes \nch\ trends for a variety of collision systems
  \bea \label{aanchx}
  n_{ch} &=& n_s (N_{part}/2) + \tilde n_h(\nu) N_{bin} 
  \\ \nonumber
  \frac{2}{N_{part}}n_{ch} &=& n_{pp} [1 + x(\nu) (\nu - 1)],
  \eea
 where $n_{pp} = n_s + n_{h}$ and $x(\nu) = n_h'(\nu) / n_{pp}$. Note that $\tilde n_h(\nu)$ is an average over all $\nu$ \nn\ encounters whereas $n_h'(\nu)$ or $x(\nu)$ applies only to the  $\nu - 1$ subsequent or secondary encounters.

For a self-consistent description the same argument should be applied to $\bar p_{t,h}(\nu)$ such that in the first \nn\ encounter the \pp\ value holds while thereafter the value may change. The \aa\ TCM for $\bar p_t(\nu)$ with \pt\ cut is then
  \bea \label{aatcmx}
 \bar p'_{t,AA}(\nu) &=&  \frac{n_s' \bar p_{t,s}' (N_{part}/2) + \tilde n_h(\nu)  \tilde p_{t,h}(\nu)N_{bin}}{n_s'(N_{part}/2) + \tilde n_h(\nu)N_{bin}} \\ \nonumber
 &\approx &  \frac{ \bar p_{t,pp}  + x(\nu)\, \bar p_{t,h}(\nu) (\nu-1)}{n_{pp}'/n_{pp} + x(\nu) \,(\nu-1)},
  \eea
where $n_{pp}'/n_{pp} \approx 0.75$ for $p_{t,cut} \approx 0.18$ GeV/c.
For a complete \aa\ $\bar p_t$ TCM description it remains to define quantities $x(\nu)$ and $\bar p_{t,h}(\nu)$.
$x(\nu)$ is defined in the second line of Eq.~(\ref{nchalice}) with $\nu_0 = 2$, $x_0 = 0.028$ and $x_1 = 0.141$ for 2.76 TeV \pbpb\ collisions. The hard component $\bar p_{t,h}(\nu)$  for the same system is defined by
\bea
\bar p_{t,h}(\nu) \hspace{-.07in} &=& \hspace{-.07in} 1.00 + 1.70 \{ 1 - \tanh[(\nu - \nu_1)/0.42] \}/2
\eea
with $\nu_1 = 1.75$, all as reported in Ref.~\cite{tommpt}.

\end{appendix}


\end{document}